\documentclass[a4paper]{article}
\usepackage[utf8]{inputenc}
\usepackage{layaureo}
\usepackage{graphicx}

\usepackage{tgpagella}
\usepackage{multirow}

\usepackage{amstext} 
\usepackage{comment}

\usepackage{amsmath}
\usepackage{tikz}
\usepackage{bm}
\usepackage[authoryear,round]{natbib}  

\usepackage{amsmath}
\usepackage{bigints}
\usepackage{float} 
\usepackage{titlesec} 
\usepackage{chngcntr}  
\usepackage{url}
\usepackage{appendix}
\usepackage{hyperref}

\title{A Poisson Factor Mixture Model for the Analysis of Linguistic Competence}

\author{
Silvia Dallari$^{1}$\thanks{Corresponding author: silvia.dallari2@unibo.it}\hspace{0.1cm},
Laura Anderlucci$^{1}$,
Nicola Grandi$^{2}$,
Angela Montanari$^{1}$\\[1ex]
$^{1}$Department of Statistical Sciences, University of Bologna\\
$^{2}$Department of Classical Philology and Italian Studies, University of Bologna
}

\date{}

\begin{document}

\maketitle
\begin{abstract}
Dependence among variables and unobserved heterogeneity are common features of multivariate count data. We propose a model-based clustering approach for this setting based on a Poisson factor mixture model, which combines a low-dimensional latent structure with a finite mixture specification. This enables simultaneous dimension reduction and clustering while providing interpretable latent representations of count profiles. The proposal originates from the UniversITA project, which collected formal written texts from a representative sample of Italian university students. Since university students play a central role in the process of linguistic change, as they are the most educated segment of the youth population, understanding variation in their writing is substantively important for both language research and higher education policy. Texts were annotated for various linguistic features, producing multivariate count data. The fitted model identifies two correlated latent dimensions of writing competence, interpretable as communicative effectiveness and grammatical-linguistic control. Incorporating educational and sociodemographic covariates further reveals distinct student profiles associated with field of study and prior educational background. Beyond this application, the proposed framework can be employed for the analysis of heterogeneous multivariate count data with complex dependence patterns, such as those arising in educational assessments, psychological and behavioral studies, and text analysis.
\\


\end{abstract}

\section{Introduction}
Multivariate count data arise in a wide range of applications, including text analysis, educational assessments, psychological and behavioral studies, but also microbiome, genomics, and ecological applications. In many of these settings, the observed counts exhibit substantial dependence across variables, are generated by heterogeneous populations, and, in some contexts, they are also sparse. These features pose important challenges for statistical modeling. Independence assumptions are often unrealistic, while highly parameterized multivariate count models can be difficult to estimate and interpret. At the same time, in many applications the inferential goal is not only to model dependence, but also to uncover low-dimensional latent structure and identify unobserved subpopulations. This motivates the development of model-based clustering methods for multivariate count data that can simultaneously account for dependence and heterogeneity in a parsimonious and interpretable way, also allowing for the inclusion of covariates. In this paper, we develop such a framework through a Poisson factor mixture model. The proposed approach combines a low-dimensional latent factor structure, which captures dependence among observed counts, with a finite mixture specification, which accommodates unobserved heterogeneity across observational units. 

The proposed framework is motivated by the analysis of the data from the UniversITA project (\url{https://site.unibo.it/univers-ita/it}; \cite{grandi}), which collected formal written texts from a nationally representative sample of Italian university students. University students occupy a particularly informative position in processes of linguistic change: they belong to younger generations, among which linguistic innovation often emerges, while also representing the most highly educated segment of that population, for whom adherence to standard norms is generally expected. Their formal writing therefore provides a useful setting for studying the interplay between educational and social background, and ongoing language change.

In the Italian context, public discussion has often framed the writing of younger generations in terms of a supposed decline in linguistic competence. Similar concerns have been documented in other national contexts, where young people's writing practices have likewise been associated with a perceived deterioration of language standards \citep{LinguistesAtterrees2023,PlewniaWitt2014,Aitchison2000,MilroyMilroy1999,Crowley2012,KahaneKahane1979}. That claim, however, is difficult to formulate and evaluate scientifically. A more meaningful empirical question concerns how the writing of contemporary university students relates to the traditional standard. Recent sociolinguistic work has emphasized that contemporary Italian is shaped by a process of restandardization, whereby the traditional literary standard coexists with a newer reference variety that has emerged through the spread of Italian as an everyday spoken and written language \citep[see, for example,][]{berruto,ballare2020,Ballare2025}. 

Despite the relevance of this topic, a comprehensive quantitative mapping of the formal writing of Italian university students has only recently become possible. The UniversITA project was designed to fill this gap by collecting original written texts together with rich background information from students enrolled across Italy. Each student was asked to write a formal text of 250 to 500 words on a common topic and to complete a detailed socio-biographical questionnaire. Subsequently, all texts were read and annotated manually and qualitatively by two expert readers with solid linguistic and metalinguistic competence and with specific training in Italian linguistic. Specifically, all forms and structures that deviate from the expected result, that is a formal text fully compliant with the rules of standard Italian, were annotated. Seven key features were identified, spanning orthography, lexicon, syntax, morphosyntax, coherence, register, and sentence structure. For each text, the number of occurrences of each feature was recorded, yielding a multivariate vector of low-frequency counts. These features are not naturally modeled as independent. Some tend to co-occur because they reflect broader underlying dimensions of writing competence or stylistic orientation, while others may vary systematically across unobserved groups of students. 

To address this structure, we formulate a Poisson factor mixture model in which the count variables are conditionally linked through a low-dimensional Gaussian latent structure, while heterogeneity is captured by a finite mixture distribution. The lower-dimensional representation is important both statistically and substantively: it supports efficient dimension reduction, facilitates interpretation, and enables the identification of latent profiles that can be meaningfully characterised in the application.

The proposed Poisson factor mixture model belongs to the broader family of generalized linear latent variable models (GLLVMs) and extends the standard Poisson GLLVM by replacing the Gaussian distribution of the latent factors with a finite mixture of multivariate Gaussian distributions. This allows dimension reduction and model-based clustering to be performed simultaneously, building on the factor mixture modeling framework of \cite{cagnoneviroli}.

Our proposal is also related to recent work on mixtures of multivariate Poisson-lognormal distributions (MPLN);
\citep[see, among others,][]{subedi2019,subedi2020,robin}. These models provide a flexible framework for dependent count data and latent heterogeneity, but they typically introduce one latent variable for each observed variable. In contrast, our approach corresponds to a Poisson-lognormal mixture with component covariance matrices of rank at most $q$, where $q$ is the number of latent factors that is smaller than the number of observed variables. Unlike conventional MPLN mixtures, the model imposes a common low-dimensional structure on both component means and covariance matrices, thereby achieving dimension reduction while retaining a latent representation suitable for clustering and interpretation. In this sense, the proposed model lies at the intersection of multivariate count modeling, latent variable methods, and model-based clustering. 

Although our motivation comes from the analysis of linguistic data, the proposed framework applies more broadly to heterogeneous multivariate count data with complex dependence structures.

The paper is structured as follows. In Section \ref{sec2} the dataset obtained after annotations is described, along with the metadata. Section \ref{sec3} introduces the proposed model, which is a mixture of generalized linear latent variable models for multivariate count data. Section \ref{sec4} addresses identifiability conditions, while Section \ref{sec5} outlines the main steps of the generalized Expectation-Maximization algorithm exploited for model estimation. Section \ref{sec6} illustrates how the model can be extended to incorporate covariates, while Section \ref{sec7} discusses the selection of the number of latent variables and clusters. In Section \ref{sec8} a thorough simulation study is conducted and Section \ref{sec9} applies the model to the project data.  Finally, conclusions and future developments are outlined in Section \ref{sec10}.

\section{The data}\label{sec2}

The considered dataset \citep{datasetuni} contains information on a sample of 2160 university students that come from different universities in Italy. After excluding observations with missing data, the final sample comprised 2137 students. For each student, the dataset includes the number of annotations received by their text. In particular, the annotated features are the following (labels in brackets):
\begin{itemize}
\item Orthography (nORT), e.g. absence or incorrect use of the apostrophe, or use of the accent with monosyllabic verb forms.
\item Linguistic register (nREG), e.g. vocabulary not adequate for the formal written context.
\item Marked sentences (nMRC), e.g. sentences which violate the natural order of constituents in a typical unmarked sentence (such as sentences with right or left dislocation, as well as cleft and pseudo-cleft sentences).
\item Lexicon (nLES), e.g. lexical poverty or excessive generality, improper vocabulary and repetitions.
\item Morphosyntax (nMFS), e.g. lack of agreement on gender and number of failures to comply with the \textit{consecutio temporum}.
\item Coherence (nCOE), e.g. illogical use of connectives, failure to explain the logical relationships that exist between the contents expressed, and contradictions.
\item Syntax (nSIN), e.g. omission of the preposition in the coordination of phrases, absolute gerunds and missing or incorrect parallelisms. 
\end{itemize}
For additional examples of these features cf. \cite{BallareGrandiPascoli2025}, p. 82-83.
 
The distribution of annotations in the sample is shown in Figure \ref{boxplot_annotations}.

The responses to the socio-biographical questionnaire mentioned above provide us with a considerable amount of metadata. The questionnaire consisted of several items, divided into four sections:
\begin{itemize}
\item Personal and family profile;
\item Linguistic and educational background;
\item Cultural consumption;
\item Attitudes and experiences related to writing. 
\end{itemize}

For example, information on respondents’ socio-economic backgrounds was collected, including gender, reading habits, family socio-economic status, family origin, and whether they were working students. Additional details were also recorded, such as the type of high-school diploma obtained, the level and field of the current degree program, and the geographical area of the university attended. This set of information makes it possible to identify systematic relationships between specific social parameters and types of texts. For instance, it is possible to detect profiles of students who display particular weaknesses in formal writing and, consequently, to design tailored teaching interventions. Similarly, one may identify degree programs in which written production is notably weak and propose curriculum revisions aimed at addressing these gaps, among other potential applications.

\section{Poisson factor mixture model for multivariate count data}\label{sec3}
The considered data are expressed as counts, as they represent the number of annotations related to each feature. Given their discrete nature and their low frequence, a suitable representation can be provided by the Poisson distribution. In order to account for the unobserved heterogeneity of the students while modeling the distribution of the features, a model-based clustering approach is considered. Specifically, we introduce a mixture of generalized linear latent variable models \citep{bartholomew} for multivariate count data, that allows to explicitly account for possible correlations among the observed variables.

In this approach, counts are modeled as Poisson distributed random variables. 
For each individual, the $p$-dimensional vector $\mathbf{y}$ of observed count variables is related with a $q$-dimensional latent vector $\mathbf{z}$ (with $q < p$) distributed according to a finite mixture of $k$ multivariate Gaussians. 
A latent allocation variable $\mathbf{s}$ ($\mathbf{s} \in R^k$) identifies the mixture component membership, with $s_i = 1$ if the observation belongs to the $i$-th component and $s_i = 0$ otherwise; therefore, the model has two latent layers and one observation layer. More formally:
\vspace{0.3cm}

\textbf{Latent layers} \\
\begin{equation}
f(\mathbf{z})=\sum_{i=1}^k \pi_i \phi_i^{(q)}(\boldsymbol{\mu}_i, \boldsymbol{\Sigma}_i)
\end{equation}
\begin{equation}
f(\mathbf{s})=\prod_{i=1}^k\pi_i^{s_i}
\end{equation}

\textbf{Observation layer} \\
\begin{equation}
f(\mathbf{y}|\mathbf{z})=\prod_{j=1}^p f(y_j|\mathbf{z})=\prod_{j=1}^p \dfrac{\omega_j(\mathbf{z})^{y_j}exp\{-\omega_j(\mathbf{z})\}}{y_j!}
\label{obslay}
\end{equation}
where $\boldsymbol{\pi}$ is the vector of mixing proportions satisfying the constraints: $\pi_i >0$ and $\sum_{i=1}^k \pi_i=1$; $\phi_i^{(q)}$ is the $q$-variate normal density of component $i$, with $\boldsymbol{\mu}_i$ and $\boldsymbol{\Sigma}_i$ being the mean vector and the covariance matrix, respectively. The relation between $E(\mathbf{y})$ and $\mathbf{z}$ can be modeled through the link function:
\begin{equation}
log(\boldsymbol{\omega}(\mathbf{z}))=\boldsymbol{\lambda}_0+\boldsymbol{\Lambda}\mathbf{z},
\label{eq:link}
\end{equation}
where $\boldsymbol{\lambda}_0$ is the vector of dimension $p\times 1$ containing the intercepts, $\boldsymbol{\Lambda}$ is the $p\times q$ factor loading matrix, and $\boldsymbol{\omega}(\mathbf{z})=(\omega_1(\mathbf{z}), \ldots, \omega_p(\mathbf{z}))$ is the $p$-dimensional vector of the Poisson random variable parameters. In the observation layer (Equation \ref{obslay}) it is assumed that the latent variables wholly explain the possible associations between the observed variables, as is typically the case in classical factor analysis. These latent variables are common to all mixture components. Observations are assumed to cluster in the latent space, which may in turn induce a clustering structure in the observed space of counts. This approach represents an extension of the models of \cite{montanariviroli} and \cite{cagnoneviroli}.


It is worth stressing that the latent layer $\mathbf{z}$ serves both to reduce dimensionality, under the assumption that $q<p$, and to model potential correlations through the set of covariance matrices $\boldsymbol{\Sigma}$. Moreover, the allocation variable $\mathbf{s}$ is used to assign samples to clusters, thus enabling the identification of a group structure within the data. Observations are assigned to the cluster with the highest posterior probability.

\subsection{Identifiability conditions}\label{sec4}

The latent factor representation in Equation \ref{eq:link} is subject to parameter indeterminacies, since different parameterizations may yield the same distribution of the observed counts. In particular, affine transformations of the latent variables can be compensated by corresponding transformations of the factor loadings and component-specific Gaussian parameters, leaving the observed-data distribution unchanged. To address these indeterminacies, we impose the following constraints:
\begin{enumerate}
\item The marginal distribution of the latent factors is standardized to have zero mean and identity covariance matrix:
\[E(\mathbf z)=\sum_{i=1}^k\pi_i\boldsymbol\mu_i=\mathbf 0,
\qquad
\operatorname{Var}(\mathbf z)=\sum_{i=1}^k\pi_i(\boldsymbol\Sigma_i+
\boldsymbol\mu_i\boldsymbol\mu_i^\top)=\mathbf I_q.
\]

\item The upper triangular elements of the factor loading matrix $\boldsymbol\Lambda$, of dimension $p\times q$, are fixed to zero, imposing $q(q-1)/2$ restrictions (in line with the proposal by \cite{joreskog} to fix $q(q-1)/2$ loadings equal to zero). This constraint addresses the rotational indeterminacy of the factor representation.
\end{enumerate}

We additionally set the first element of the intercept vector to zero ($\lambda_{10}=0$). This parameterization restriction improves the estimate stability, but it does not represent a general identifiability requirement.


\subsection{Model Estimation}\label{sec5}
The complete density can be written as:
\begin{equation}
f(\mathbf{y}, \mathbf{z}, \mathbf{s})=f(\mathbf{y}|\mathbf{z}) f(\mathbf{z}|\mathbf{s})f(\mathbf{s}),
\end{equation}
and, by taking the logarithm, it follows that:
\begin{equation}
log \,f(\mathbf{y}, \mathbf{z}, \mathbf{s})=log \,f(\mathbf{y}|\mathbf{z})+ log\, f(\mathbf{z}|\mathbf{s})+log\,f(\mathbf{s}).
\label{scomp}
\end{equation}
The model can be estimated by a generalized Expectation-Maximization (EM) algorithm \citep{dempster,gem}.

Considering a sample of size $n$ and denoting the corresponding variables with $\mathbf{y}=(\mathbf{y}_1, \ldots,\mathbf{y}_l, \ldots,\mathbf{y}_n)$, $\mathbf{z}=(\mathbf{z}_1, \ldots,\mathbf{z}_l, \ldots,\mathbf{z}_n)$, $\mathbf{s}=(\mathbf{s}_1, \ldots,\mathbf{s}_l, \ldots,\mathbf{s}_n)$ and the whole set of parameters with $\boldsymbol{\theta}=(\boldsymbol{\lambda}_0,\boldsymbol{\Lambda},\boldsymbol{\pi}, \boldsymbol{\mu}, \boldsymbol{\Sigma})$, where $\boldsymbol{\pi}=(\pi_1, \ldots,\pi_k)$, $\boldsymbol{\mu}=(\boldsymbol{\mu}_1, \ldots,\boldsymbol{\mu}_k)$ and $\boldsymbol{\Sigma}=(\boldsymbol{\Sigma}_1, \ldots,\boldsymbol{\Sigma}_k)$, model estimation can be performed via an EM algorithm that consists in solving:

$$\underset{\theta}{argmax}\,E_{\mathbf{z},\mathbf{s}|\mathbf{y},\boldsymbol{\theta}'}\biggl[\sum_{l=1}^n log f(\mathbf{y}_l, \mathbf{z}_l,\mathbf{s}_l;\boldsymbol{\theta})\biggr]. $$
Given Equation \ref{scomp}, the maximization simplifies in:

\begin{equation}
\begin{aligned}
\underset{\boldsymbol{\theta}}{argmax}\;
E_{\mathbf{z},\mathbf{s}\mid\mathbf{y},\boldsymbol{\theta}'}
\!\left[\log f(\mathbf{y}, \mathbf{z},\mathbf{s};\boldsymbol{\theta})\right]
&=
\underset{\boldsymbol{\theta}}{argmax}\;
E_{\mathbf{z},\mathbf{s}\mid\mathbf{y},\boldsymbol{\theta}'}
\!\left[
\log f(\mathbf{y}\mid\mathbf{z};\boldsymbol{\theta}) \right. \\
&\qquad\left. + \log f(\mathbf{z}\mid\mathbf{s};\boldsymbol{\theta})
+ \log f(\mathbf{s};\boldsymbol{\theta})
\right].
\end{aligned}
\label{eqem}
\end{equation}
The integrals in the E-step of the algorithm cannot be solved analytically and, thus, they need to be approximated numerically. To do that, we adopt a weighted sum over a finite number of
points with weights given by Gauss-Hermite quadrature points \citep[see][]{straud}. Gauss-Hermite quadrature points allow to have estimates in closed form, with the only exception of the estimator for ($\boldsymbol{\lambda}_0, \boldsymbol{\Lambda}$) that needs to be derived with Newton-Raphson procedures, leading to a generalized EM algorithm \citep{gem}.\\

The steps of the algorithm can be summarized as follows:
\begin{enumerate}
\item Choose the starting values of the parameters $\boldsymbol{\theta'}=(\tilde{\boldsymbol{\Lambda}}', \boldsymbol{\pi'}, \boldsymbol{\mu'}, \boldsymbol{\Sigma'})$, where $\tilde{\boldsymbol{\Lambda}}=(\boldsymbol{\lambda}_0,\boldsymbol{\Lambda})$.

\item Find:
\begin{itemize}
\item $\tilde{\boldsymbol{\Lambda}}$ that maximizes $E_{\mathbf{z},\mathbf{s}|\mathbf{y},\boldsymbol{\theta'}}[log f(\mathbf{y}|\mathbf{z};\mathbf{\boldsymbol{\theta}})];$
\item $\boldsymbol{\mu}$, $\boldsymbol{\Sigma}$ maximizing $E_{\mathbf{z},\mathbf{s}|\mathbf{y},\boldsymbol{\theta'}}[log f(\mathbf{z}|\mathbf{s};\boldsymbol{\theta})];$
\item $\boldsymbol{\pi}$ that maximizes $E_{\mathbf{z},\mathbf{s}|\mathbf{y},\boldsymbol{\theta'}} [log f(\mathbf{s;\boldsymbol{\theta}})].$
\end{itemize}	
\item Set $\boldsymbol{\theta'}=\boldsymbol{\theta}$.

\item If convergence is not obtained return to Step 2. Convergence is achieved when the relative change in the observed log-likelihood between two subsequent steps is smaller than a fixed $\epsilon$.
\end{enumerate}

As is generally the case with the EM algorithm, the proposed procedure is sensitive to initialization. To prevent it from getting stuck in local maxima, we recommend to initialize the factor loadings with the solution obtained from the classical factor analysis using maximum likelihood or principal factors methods (correcting the results according to the identifiability conditions), and the clustering allocation with the solution of another clustering method, e.g., the Ward's method \citep{ward}. If a random initialization is chosen, a multi-start strategy is advisable.

The detailed steps of the algorithm are reported in Appendix \ref{sec11}.

%

\subsection{Extension with covariates}\label{sec6}
If available, covariates can also be added to the model. In particular, they may be introduced both in the mean of the latent variables and in the a priori probability of cluster membership. In this work the second option was adopted, as the goal of including metadata was to enhance cluster interpretability and to promote cluster formation. The chosen variables can enter the mixture weights following a multinomial logit regression as proposed by \cite{fokoue}. Supposing that for each unit we observe $m$ covariates, and denoting by $\mathbf{x}$ the resulting $(m+1)-$dimensional vector, then:
\begin{equation}
\pi_i(\mathbf{x})=\dfrac{exp(\boldsymbol{\eta}_i^T\mathbf{x})}{1+\sum_{i'=1}^{k-1}exp(\boldsymbol{\eta}_{i'}^T\mathbf{x})},
\end{equation}
where $\boldsymbol{\eta}_i=(\eta_{0i},\eta_{1i}, \ldots,\eta_{mi})$ is the vector of regression coefficients, with $i=1, \ldots,k$ and $\boldsymbol{\eta}_k=0$ for identifiability reasons. The gradient does not offer an explicit solution for $\boldsymbol{\eta}_i$, $i=1, \ldots,k$, thus nonlinear optimization methods need to be used to estimate them. In this work a Newton-Raphson algorithm is adopted. 

If covariates are added to the model, the EM procedure is the same as for the case without covariates, with the only exception of the mixture weights that need to be optimized as just mentioned. 

\subsection{Model selection}\label{sec7}
In the proposed mixture model, two quantities must be specified: the number of factors $q$ and the number of groups $k$.
To determine suitable values, we fit the model across a range of $(q,k)$ combinations and select the optimal pair according to established information criteria. Following approaches used in \cite{MCLACHLAN}, \cite{Baek}, \cite{cagnoneviroli} and \cite{Cagnone2014}, we rely on the Bayesian Information Criterion \citep[BIC;][]{bic} and the Akaike Information Criterion \citep[AIC;][]{Akaike}. Both criteria aim to identify the number of groups that achieves the best balance between goodness of fit and model complexity, with BIC generally favoring more parsimonious models compared to AIC.

The former is defined as:
\begin{equation}
BIC=-2\,log(L)+h\, log(n),
\end{equation}
where $log(L)$ is the log-likelihood, $h$ is the number of model parameters and $n$ is the number of observations, while the latter, i.e. AIC, as:
\begin{equation}
AIC=-2\,log(L)+2h.
\end{equation}

In the proposed mixture model the number of parameters $h$ is $p\times(q+1)-(q\times(q-1)/2+1)+(q\times(q+1)/2)\times(k-1)+q\times(k-1)+k-1$ if no covariate is added to the mixture weights, and $p\times(q+1)-(q\times(q-1)/2+1)+(q\times(q+1)/2)\times(k-1)+q\times(k-1)+(k-1)\times(m+1)$ if there are covariates.

In the simulation study presented below, we evaluate the performance of both criteria in correctly identifying the true number of factors and clusters.

\section{Empirical Results}
\subsection{Simulation study}\label{sec8}
The proposed model has been tested on simulated data before being applied to real data. Different scenarios have been studied by considering different numbers of factors ($q=1,2$), groups ($k=2,3,6$), samples ($n=300$, $n=2000$) and variables ($p=10$, $p=50$). 
In each experiment the factor loadings in $\boldsymbol{\Lambda}$ 
are generated in such a way that a quasi simple structure \citep{Thurstone} is obtained. 
In all the experiments, condition 2) of Section \ref{sec4} is applied to the loadings during data generation. The mean and the covariance matrix of the factors have been chosen so that in the latent space the clusters are quite well separated and the factors are standardized as required by constraint 1 (Section \ref{sec4}). For illustration, when $q=2$ and $k=3$, the means and covariance matrices of the latent factors, along with the mixture weights, are specified as follows:

$$
\boldsymbol{\mu}_1=\left[
\begin{array}{cc}
1.200 & 0.760  \\
\end{array}
\right]
\qquad
\boldsymbol{\mu}_2=\left[
\begin{array}{cc}
-1.190 & 0.770  \\
\end{array}
\right]
\qquad
\boldsymbol{\mu}_3=\left[
\begin{array}{cc}
-0.0075 & -1.148  \\
\end{array}
\right]
$$

$$
\boldsymbol{\Sigma}_1=\left[
\begin{array}{cc}
0.170 & 0.080  \\
0.080&0.140\\
\end{array}
\right]
\qquad
\boldsymbol{\Sigma}_2=\left[
\begin{array}{cc}
\phantom{-}0.160 & -0.080  \\
-0.080 & \phantom{-}0.120\\
\end{array}
\right]
\qquad
\boldsymbol{\Sigma}_3=\left[
\begin{array}{cc}
\phantom{-}0.110 & -0.005  \\
-0.005 & \phantom{-}0.110\\
\end{array}
\right]
$$

$$\boldsymbol{\pi}=[0.3,0.3,0.4] $$

\begin{figure}[b]
\begin{center}
\includegraphics[width=\linewidth]{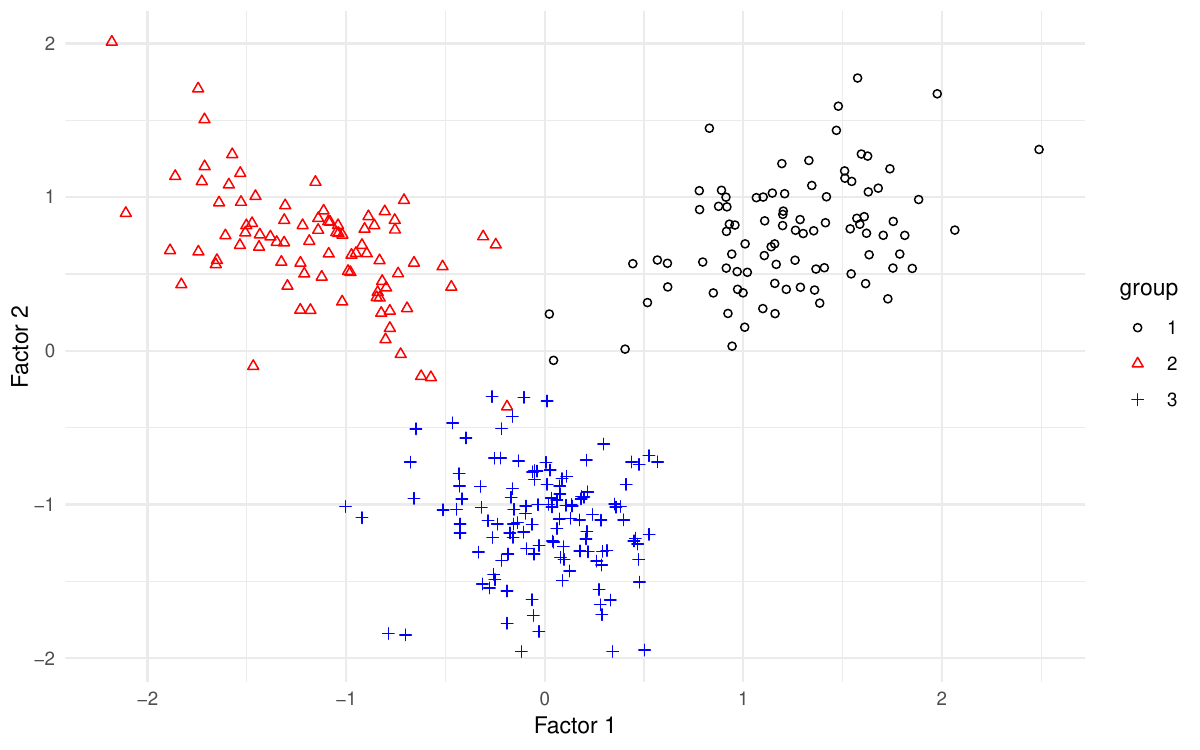}
\caption{Scatterplot of the factor scores distinguished by group for one of the 100 replicates of the experiment with $q=2$, $k=3$, $n=300$ and $p=50$.}
\label{figsim1}
\end{center}
\end{figure}

The scatterplot of the factor scores for one of the 100 replicates of this experiment (with $n=300$ and $p=50$) is reported in Figure~\ref{figsim1}.

The ability of BIC and AIC to select the optimal combination of the number of latent factors $q$ and clusters $k$ was assessed across multiple simulation settings, each replicated 100 times. Since the intercepts and cluster means were randomly initialized, five different initialization seeds were used for each simulation replicate. 
For every configuration, the solution achieving the best BIC or AIC value was retained.

 \normalsize

\begin{table}
\caption{Capability of BIC and AIC of recovering the true number of clusters and latent factors. Displayed results refer to $n=2000$ and $p=10$. For each setting, bold identifies the true model specification and the corresponding percentage of times such model was selected by each criterion.}
\setlength{\tabcolsep}{4pt}
\begin{center}
\begin{tabular}{lccc|ccc|lccc|ccc}

\hline
& \multicolumn{3}{c}{BIC}& \multicolumn{3}{c|}{AIC} & & \multicolumn{3}{c}{BIC}& \multicolumn{3}{c}{AIC} \\
\hline
%

&\multicolumn{6}{c|}{$k=2, q=1$}&&\multicolumn{6}{c}{$k=2, q=2$}\\

& $\boldsymbol{q=1}$ &  $q=2$  & $q=3$
& $\boldsymbol{q=1}$ &  $q=2$  & $q=3$ & &$q=1$ &  $\boldsymbol{q=2}$  & $q=3$
& $q=1$ &  $\boldsymbol{q=2}$  & $q=3$ \\
\hline
$k$=1 &0\% &0\%&0\% &0\%&0\%&0\%&$k$=1 &0\% &0\%&0\% &0\%&0\%&0\%\\
$\boldsymbol{k}$=\textbf{2} &$\mathbf{100}$\% &0\% & 0\%&$\mathbf{100}$\%&0\%&0\%&
$\boldsymbol{k}$=\textbf{2} &0\% &$\mathbf{100}$\% & 0\%&0\%&$\mathbf{89}$\%&7\%\\
k=3 &0\% &0\%&0\%&0\%&0\%&0\% &$k$=3 &0\% &0\%&0\%&0\%&4\%&0\%\\
$k$=4 &0\% &0\%&0\%&0\%&0\%&0\%& $k$=4 &0\% &0\%&0\%&0\%&0\%&0\%\\

\hline
%
%

%
%

&\multicolumn{6}{c|}{$k=3, q=1$}&&\multicolumn{6}{c}{$k=3, q=2$}\\

& $\boldsymbol{q=1}$ &  $q=2$  & $q=3$
& $\boldsymbol{q=1}$ &  $q=2$  & $q=3$ &&$q=1$ &  $\boldsymbol{q=2}$  & $q=3$
& $q=1$ &  $\boldsymbol{q=2}$  & $q=3$  \\
\hline
$k$=1 &0\% &0\%&0\% &0\%&0\%&0\% & $k$=1 &0\% &0\%&0\% &0\%&0\%&0\%\\
$k$=2 &0\% &0\% & 0\%&0\%&0\%&0\%& $k$=2 &0\% &0\% & 0\%&0\%&0\%&0\%\\
$\boldsymbol{k}$=\textbf{3} &$\mathbf{100}$\% &0\%&0\%&$\mathbf{97}$\%&3\%&0\%&$\boldsymbol{k}$=\textbf{3} &0\% &$\mathbf{100}$\%&0\%&0\%&$\mathbf{99}$\%&1\%\\
$k$=4 &0\% &0\%&0\%&0\%&0\%&0\%&$k$=4 &0\% &0\%&0\%&0\%&0\%&0\%\\

\hline

%
%
%
%

&\multicolumn{6}{c|}{$k=6, q=1$}&&\multicolumn{6}{c}{$k=6, q=2$}\\

& $\boldsymbol{q=1}$ &  $q=2$  & $q=3$
& $\boldsymbol{q=1}$ &  $q=2$  & $q=3$  && $q=1$ &  $\boldsymbol{q=2}$  & $q=3$
& $q=1$ & $\boldsymbol{q=2}$  & $q=3$ \\
\hline
$k$=1 &0\% &0\%&0\% &0\%&0\%&0\%&$k$=1 &0\% &0\%&0\% &0\%&0\%&0\%\\
$k$=2 &41\% &0\% & 0\%&0\%&0\%&0\%&$k$=2 &0\% &0\% & 0\%&0\%&0\%&0\%\\
$k$=3 &49\% &0\%&0\%&56\%&2\%&0\%&$k$=3 &0\% &83\%&0\%&0\%&1\%&0\%\\
$k$=4 &10\% &0\%&0\%&31\%&4\%&0\%&$k$=4 &0\% &17\%&0\%&0\%&36\%&0\%\\
$k$=5 &0\% &0\%&0\%&6\%&0\%&0\%&$k$=5 &0\% &0\%&0\%&0\%&12\%&0\%\\
$\boldsymbol{k}$=\textbf{6} &$\mathbf{0}$\% &0\%&0\%&$\mathbf{1}$\%&0\%&0\%&$\boldsymbol{k}$=\textbf{6} &0\% &$\mathbf{0}$\%&0\%&0\%&$\mathbf{50}$\%&0\%\\
$k$=7 &0\% &0\%&0\%&0\%&0\%&0\%&$k$=7 &0\% &0\%&0\%&0\%&1\%&0\%\\

\hline
%
%
%
\end{tabular}
\end{center}
\label{table1gllvm}
\end{table}

\normalsize


The results, summarized in Table \ref{table1gllvm}, indicate that, when $k=2$ or $k=3$, both BIC and AIC generally select the correct number of $q$ and $k$. When $k=6$ and $q=1$, both methods underestimate the number of clusters, as it is difficult for data generated from a single latent variable to give rise to six well-separated components in the observed space. An inspection of the resulting partition, reported in Figure \ref{figpcak6q1} of Appendix \ref{sec12}, confirms that the clusters are not clearly distinct, which explains why both AIC and BIC tend to underestimate the true number. When $k=6$ and $q=2$, the six clusters are well separated in the latent space, whereas this separation is not preserved in the observed space (see Figure \ref{figpcak6q2} of Appendix \ref{sec12}). In this setting, AIC is able to correctly identify the six clusters in most cases, while BIC tends to underestimate their number.
Additional scenarios are presented in Table \ref{table1gllvm_appendix} of Appendix \ref{sec12}, where it can be seen that, in most cases, BIC and AIC correctly select the number of groups and latent variables. Even when this is not the case - for example, for the combination $k=2,q=2,n=2000,p=50$, where both criteria select three latent variables instead of two - Table \ref{tableariscelte} in Appendix \ref{sec12} shows that the clustering performance obtained with the models selected by the two criteria remains very good. 

\begin{figure}[tb]
\begin{center}
\includegraphics[width=\linewidth]{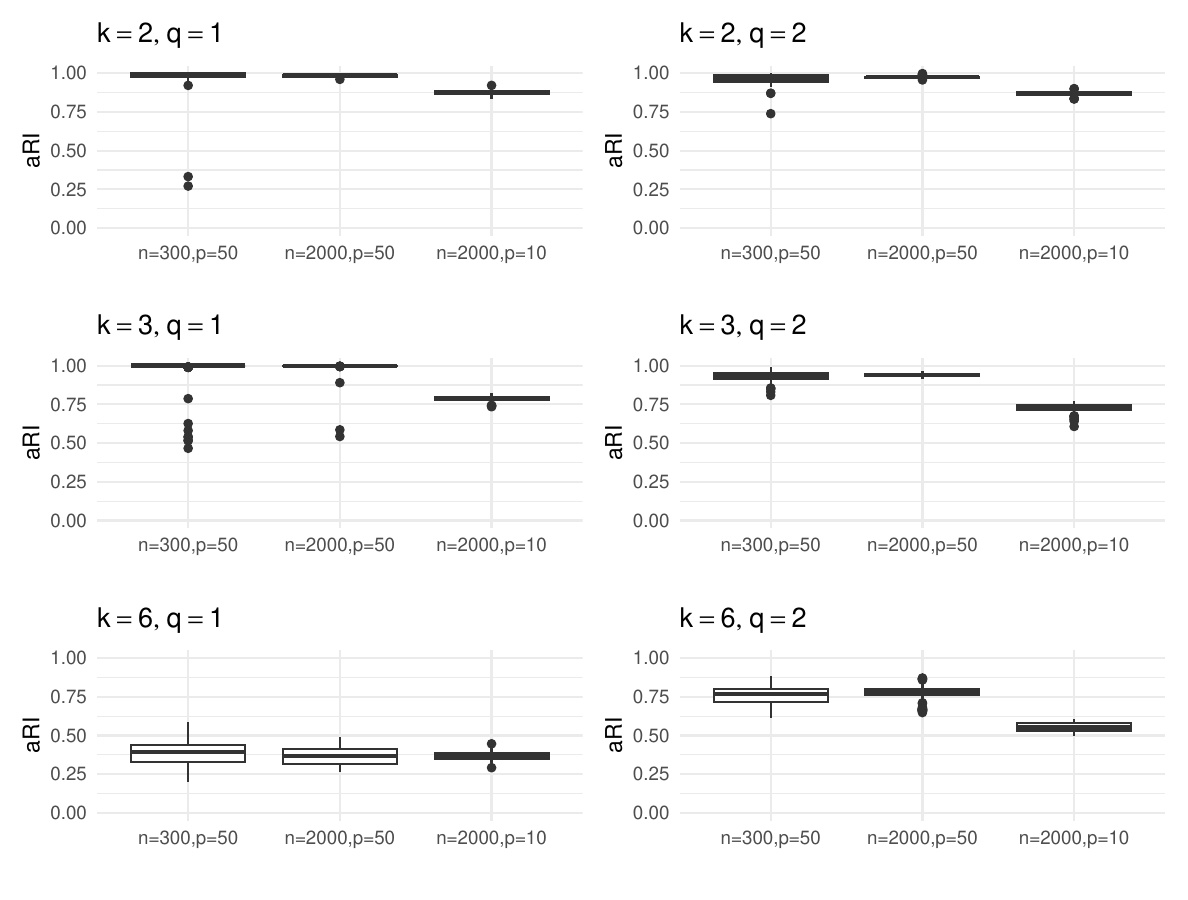
}
\end{center}
\caption{Ability of recovering the true clustering structure (with known $q$ and $k$) for the different simulation designs considered.}
\label{figboxplot}
\end{figure}

In the same simulation scenarios, clustering accuracy has been evaluated assuming that the true values of $q$ and $k$ are known. Performance has been measured using the adjusted Rand Index (aRI) over 100 replicates. For each replicate, ten different initialization seeds have been considered and the best solution according to the log-likelihood is selected. The results, reported in Figure \ref{figboxplot}, show that the model consistently achieves strong clustering performance. The only exception occurs in the scenario with $k=6$ and $q=1$, where the performance is poor due to the lack of clear separation between groups, as previously discussed and illustrated in Figure \ref{figpcak6q1}.
Notably, model performance improves as the number of observed variables $p$ increases, suggesting that a richer information set enhances the model's ability to correctly detect the clustering structure.

%
%

The accuracy of the factor loading estimates was also examined. Table \ref{table4gllvm} reports the estimated intercepts and factor loadings for the simulation design with $q=2$, $p=10$, $k=3$, and $n=2000$. Since 100 replications were performed, the table shows the average estimates, with corresponding standard deviations in parentheses. The results indicate that the estimated entries for the intercept and factor loadings are very similar to the true ones. More precise estimates could be obtained by increasing the number of quadrature points (set to eight in the present study).

\begin{table}[tb]

\caption{Estimation accuracy of the true intercepts and factor loadings for the simulation setting with $q=2, p=10, k=3, n=2000$. The average estimates are reported with the corresponding standard deviations in brackets.}

    \begin{center}
        \begin{tabular}{ccc|ccc}
\hline
\multicolumn{3}{c|}{True}&  \multicolumn{3}{c}{Estimates}  \\

$\lambda_0$ & $\lambda_1$ & $\lambda_2$ &  $\hat{\lambda}_0$ & $\hat{\lambda}_1$ & $\hat{\lambda}_2$\\
\hline
\phantom{-}0.00 & \phantom{-}0.76 & \phantom{-}0.00  & \phantom{-}0.00 (0.00) & \phantom{-}0.77 (0.03)& \phantom{-}0.00 (0.00)\\
\phantom{-}0.62 & \phantom{-}0.75 & -0.49 &\phantom{-}0.61 (0.03)& \phantom{-}0.77 (0.04)& -0.49 (0.03)\\
-0.23&\phantom{-}0.77&-0.44&-0.23 (0.04)& \phantom{-}0.78 (0.04)&-0.43 (0.04)\\
-0.34&\phantom{-}0.78&-0.45&-0.35 (0.04)&\phantom{-}0.78 (0.04)&-0.45 (0.04)\\
\phantom{-}0.20&\phantom{-}0.93&-0.38&\phantom{-}0.20 (0.03)&\phantom{-}0.94 (0.04)&-0.38 (0.04)\\
\phantom{-}0.21&-0.09&\phantom{-}0.90&\phantom{-}0.20 (0.04)&-0.08 (0.04)&\phantom{-}0.91 (0.03)\\
-0.75&-0.44&\phantom{-}0.80&-0.76 (0.04)&-0.44 (0.04)&\phantom{-}0.82 (0.05)\\
-0.41&-0.15&\phantom{-}0.96&-0.42 (0.05)&-0.14 (0.05)&\phantom{-}0.98 (0.05)\\
\phantom{-}0.16&-0.05&\phantom{-}0.78&\phantom{-}0.15 (0.03)&-0.05 (0.04)&\phantom{-}0.80 (0.03)\\
\phantom{-}0.26&-0.36&\phantom{-}0.88&\phantom{-}0.26 (0.03)&-0.36 (0.05)&\phantom{-}0.90 (0.04)\\

\hline
\end{tabular}
\end{center}
\label{table4gllvm}
\end{table}

\subsection{Real Data Analysis}\label{sec9}
Following \cite{dallari}, the model is first applied to the annotation dataset with no inclusion of covariates, taking into account different values of $q$ and $k$. 

In order to define a reasonable range for the number of factors,  as a rule of thumb we employed the Ledermann's condition \citep{lederman}.  Originally derived for classical factor analysis with a diagonal uniqueness matrix, the bound is here used solely to restrict the model search. We apply the inequality by considering the counts as states of a continuous variable; this heuristic criterion is not to be interpreted as an identifiability condition nor as an upper bound on the latent dimension. 
 Accordingly, the maximum number of latent variables $q$ will be set to 3. Hovewer, since  the model goodness of fit improvement for $q=3$ is negligible with respect to the one obtained with $q=2$, the possible combinations given by $q=1,2$ and $k=1,2,3,4,5$ have been examined. 
In this analysis, model selection is guided by the Akaike Information Criterion, prioritizing the identification of potential student clusters over model parsimony. According to AIC, the best-fitting model is the one with two latent factors and one mixture component.

An oblique rotation \citep{bartholomew} using the \texttt{oblimin} function from the \texttt{GPArotation} \texttt{R} package \citep{gparotation} is performed to allow for possible correlations between the latent factors. 
The rotated factor loadings are reported in Table~\ref{tab:table1}, and the corresponding correlation matrix of the rotated latent factors is presented in Table~\ref{tab:table2}.

\begin{table}[tb]
\caption{Loading estimates obtained after oblimin rotation for the model without covariates. Only loadings whose absolute value is greater than 0.2 are printed.}
\renewcommand{\arraystretch}{1.3}
    \begin{center}
        \begin{tabular}{llccc}
\hline 
 &\,\,& $\hat{\boldsymbol{\lambda}}_1$ & \,\,&$\hat{\boldsymbol{\lambda}}_2$\\
\hline
nCOE &\,\, & & &0.53  \\
nLES&\,\, & & &0.71  \\
nMFS&\,\, & & &0.56  \\
nMRC&\,\, &0.71 & &  \\
nORT&\,\, &0.66 & &0.41  \\
nREG&\,\, &1.34 & &  \\
nSIN&\,\, & & &0.45  \\
\hline
\end{tabular}
\end{center}

\label{tab:table1}
\end{table}

Table~\ref{tab:table1} provides useful insights into the interpretation of the two latent factors. Indeed, the high loadings on the first factor for the features referring to linguistic register and marked sentences indicate that this factor differentiates students who make inappropriate choices with respect to the context from those who do not. It can be interpreted as a \emph{communicative competence}. In contrast, the second latent factor is primarily associated with morphosyntax, lexicon, syntax and coherence. This suggests that the second factor captures difficulties in structuring a complex formal text of the type required for the task. It can be defined as \emph{linguistic competence}, marked by a distinctly grammatical emphasis. Orthography loads on both factors, as it relates to stylistic aspects relevant to both dimensions.

\begin{table}[tb]
\caption{Correlation matrix of the rotated factors obtained with the 
model without covariates.}
\renewcommand{\arraystretch}{1.2}
    \begin{center}
        \begin{tabular}{lccc}

\hline
 & $\hat{\boldsymbol{\lambda}}_1$ & &$\hat{\boldsymbol{\lambda}}_2$\\
\hline
 $\hat{\boldsymbol{\lambda}}_1$ &1.00 & &0.37  \\
$\hat{\boldsymbol{\lambda}}_2$ &0.37 & &1.00  \\
\hline
\end{tabular}
\end{center}
\label{tab:table2}
\end{table}

Table~\ref{tab:table2} shows that the two factors are correlated. The results obtained without covariates therefore suggest the presence of two distinct, yet related, dimensions of annotation: one associated with inappropriate linguistic choices and the other reflecting difficulties in structuring a formal text.\\

To further characterize the students and investigate potential underlying group structures, additional information can be incorporated as covariates in the mixture weights. Based on their relevance in the linguistic literature, the following covariates were considered:
\begin{enumerate}
\item[1)] \textit{Family origin} (Italy/Mixed/Abroad);
\item[2)] \textit{High-school diploma} (Lyceum/Professional/Technical); 
\item[3)] \textit{Campus} (South/Central/North); 
\item[4)] \textit{Gender} (Female/Male/Other);
\item[5)] \textit{Study Area} (Healthcare/Scientific/Social/Humanities);
\item[6)] \textit{Worker} (No/Yes);
\item[7)] \textit{Social Class} (Low/Middle/High/Don't know-don't answer);
\item[8)] \textit{Books read} (None/Less than five/Between five and ten/More than ten).
\end{enumerate}


In the results that follow, the category ``Don't know-don't answer'' is labeled as ``NA''. Dummy variables have been created for all covariates except for the number of books read, where the midpoint of each category has been considered in the analysis; for the open-ended category (“More than 10”), a value of 15 was assumed. Considering one level as reference category for each categorical covariate, a total of 16 variables have been included in the mixture weights. Taking into account all combinations of $q=1,2$; $k=1, \ldots,5$ and 100 different initialization seeds for each configuration, AIC identifies the optimal solution as the one with $q=2$ and $k=3$. Group 1 includes 56 students, group 2 comprises 881 individuals and group 3 the remaining ones (1200). 





The factor loadings of the two latent factors obtained after oblimin rotation are reported in Table~\ref{tab:table3}, and the corresponding correlation matrix is shown in Table~\ref{tab:table4}.
\begin{table}[b]
\caption{Loading estimates obtained after oblimin rotation for the model with covariates. Only loadings whose absolute value is greater than 0.2 are printed.}
\renewcommand{\arraystretch}{1.3}
    \begin{center}
        \begin{tabular}{llccc}
\hline 
 &\,\,& $\hat{\boldsymbol{\lambda}}_1$ & \,\,&$\hat{\boldsymbol{\lambda}}_2$\\
\hline
nCOE &\,\, & & &0.60  \\
nLES&\,\, & & &0.80  \\
nMFS&\,\, & & &0.63  \\
nMRC&\,\, &0.77 & &  \\
nORT&\,\, &0.66 & &0.52  \\
nREG&\,\, &1.45 & &  \\
nSIN&\,\, & & &0.50  \\
\hline
\end{tabular}
\end{center}

\label{tab:table3}
\end{table}

\begin{table}[tb]
\caption{Correlation matrix of the rotated factors obtained with the 
model with covariates.}
\renewcommand{\arraystretch}{1.2}
    \begin{center}
        \begin{tabular}{lccc}

\hline

 & $\hat{\boldsymbol{\lambda}}_1$ & &$\hat{\boldsymbol{\lambda}}_2$\\
\hline
$\hat{\boldsymbol{\lambda}}_1$ &1.00 & &0.40  \\
$\hat{\boldsymbol{\lambda}}_2$ &0.40 & &1.00  \\
\hline
\end{tabular}
\end{center}
\label{tab:table4}
\end{table}


The factor loading results and the correlation structure are similar to the ones obtained without covariates in Table~\ref{tab:table1} and ~\ref{tab:table2}. This conclusion is coherent with the model that assumes a unique latent space inside which the observations are clustered.

\begin{figure}[b]
\begin{center}
\includegraphics[width=\linewidth]{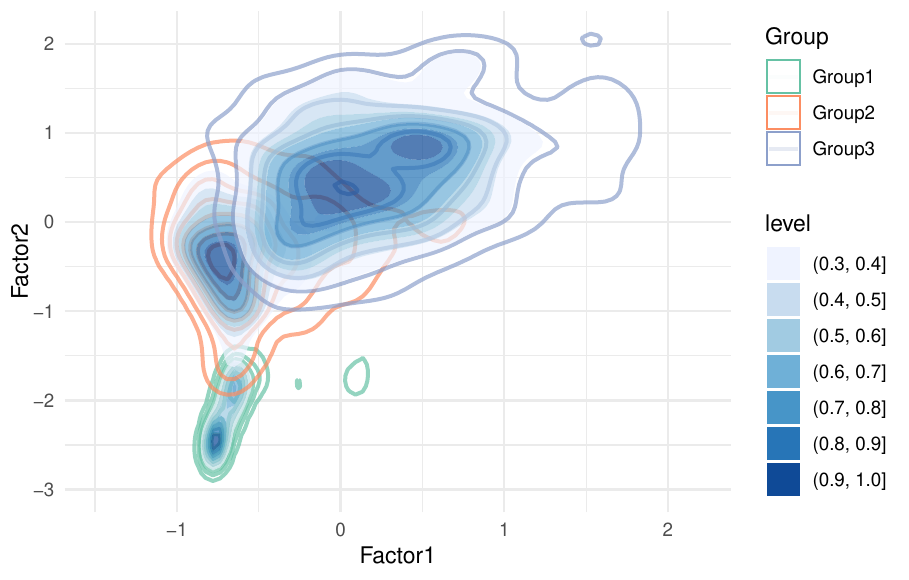}
\end{center}
\caption{Contours of the 2D kernel densities of the three groups obtained for the Univers-ITA dataset with covariates. On both dimensions lower values represent better results.}
\label{figuova}
\end{figure}

The 2D kernel densities of the estimated factor scores, distinguished by the three obtained groups, are displayed as contours in Figure \ref{figuova}. The first group (in green) shows low scores on both latent factors, representing the group of students who have received very few annotations of either type. The second group, i.e., the orange one, reflects students who have received few annotations, but have larger scores in the second factor, thus showing more difficulties in structuring a complex formal text compared to those in the first group. The third group includes the remaining individuals, i.e., those who received more annotations than the students in the other two groups on both latent dimensions. An analysis of the total word count revealed that, among the three groups, Group 1 writes, on average, fewer words, while Group 3 writes the most. Although one might suspect that the lower number of annotations received by Group 1 is simply a consequence of producing shorter texts, an examination of the ratio between annotations and total words written confirms that Group 1 still receives the fewest annotations per word, while Group 3 receives the most. 

\begin{figure}[tb]
\begin{center}
\includegraphics[width=\linewidth]{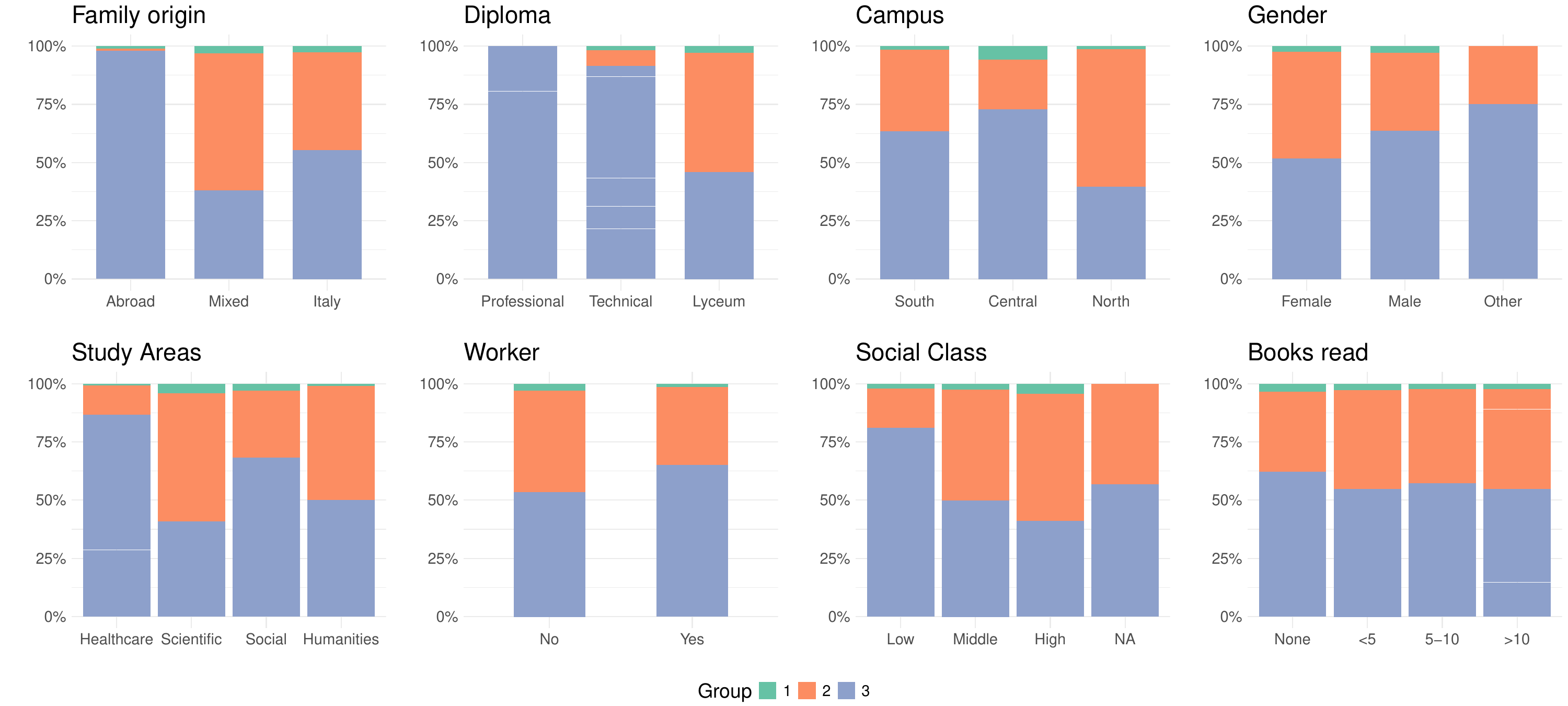}

\end{center}
\caption{Percentage of students belonging to one of the three groups for each category of the covariates considered for the Univers-ITA dataset.}
\label{figcovuni}
\end{figure}

Figure \ref{figcovuni} reports the composition of each covariate category across the identified clusters. 

The barplots show that students with family origins from abroad are mostly concentrated in cluster 3, whereas those with Italian or mixed origins display a lower proportion in this group. 

Regarding the type of high-school diploma, over 50\% of students with a lyceum diploma ended up in one of the two groups that received very few annotations (groups 1 and 2). This outcome is not surprising, as some lyceum tracks include Latin as a subject, which enhances greater linguistic awareness and may reduce the probability of errors. Conversely, nearly all the students with a professional or technical diploma belong to the third group. These diplomas emphasize practical and technical skills and typically provide less exposure to subjects that directly support linguistic development, which may contribute to the higher number of annotations observed.

As for campus location, most students from northern campuses are assigned to cluster 2. Central campuses show a higher proportion of students in group 3 but, at the same time, include the majority of individuals in group 1 (35 out of 56).

Gender and working-student status also reveal notable patterns: a smaller percentage of females and non-working students fall into group 3 compared to males and students who work. 

The number of books read appears to have minimal impact on group composition. In contrast, differences across fields of study and social class are more pronounced. Indeed, the majority of people who study in healthcare ended up in the group that received more annotations (group 3), followed by those who study in social areas. Differently, roughly half of the individuals studying humanities subjects and the majority of students from scientific areas ended up either in group 1 or in group 2, hence receiving very few annotations. Regarding the social class, it seems that the proportion of students in groups 1 and 2 is higher among those from middle or upper classes.

\section{Discussion}\label{sec10}
In this paper a mixture of Poisson generalized linear latent variable models for multivariate count data, which can be seen as an alternative to the MPLN mixtures of \cite{subedi2019}, \cite{subedi2020} and \cite{robin}, has been proposed. Differently from those proposals, the mixture presented in this work considers a number of latent variables smaller than the number of observed ones. Because the MPLN mixture does not perform dimension reduction, the two approaches are not directly comparable in terms of interpretability.

A comprehensive simulation study demonstrated that the proposed model achieves good clustering accuracy and yields precise parameter estimates. Both the Bayesian Information Criterion and the Akaike Information Criterion proved effective in selecting the number of groups and latent variables. The choice between the two depends on the desired level of parsimony: BIC is recommended when a more parsimonious model is preferred, whereas AIC is more suitable when allowing for greater model complexity, as it tends to select richer structures.

As a possible future development, alternative estimation procedures — such as the Variational EM algorithm — could be explored to reduce computational costs and, among other advantages, to allow for a larger number of latent variables to be considered.
Lastly, the algorithm shows sensitivity to starting points, a common issue with EM-based methods, which can be mitigated by choosing the most appropriate initialization.


In the real data analysis, the fact that the seven types of annotation are grouped coherently into two areas — Factor 1, related to the ability to make appropriate register choices, for which lower factor scores correspond to stronger communicative abilities, and Factor 2, related to the ability to construct fully grammatically correct texts, for which lower factor scores indicate higher levels of linguistic and grammatical competence — perfectly reproduces the structure of speakers’ competence. 
Speakers’ competence is in fact divided into two ‘sub-competences’. 
One is communicative competence, which concerns the consistency between linguistic structures and the context in which they are used and allows speakers to make judgments about appropriateness with respect to the communicative situation. The other, which we can define as linguistic competence, has a more strictly grammatical dimension and allows speakers to make judgments about the grammaticality of linguistic structures \citep[cf., among others,][]{berruto}. It is well known, in fact, that constructing a perfectly grammatical sentence is no guarantee of successful communication. A sentence that is perfectly formed from a grammatical point of view may be completely inappropriate to the situation. Mastering a language therefore means making judgments about both grammaticality and appropriateness.
In Italy, language education in schools is heavily biased towards grammatical competence and neglects communicative competence. Public debate is also affected by this distortion and is heavily focused on grammatical errors, as demonstrated by the so-called ‘Letter of the 600’, which inspired the Univers-ITA project (\url{https://gruppodifirenze.blogspot.com/2017/02/contro-il-declino-dellitaliano-scuola.html}).
This disconnection between language education in schools and the actual use of language has been repeatedly pointed out, starting with the ‘Dieci tesi per l’educazione linguistica democratica’ (\url{https://giscel.it/dieci-tesi-per-leducazione-linguistica-democratica/}; cf., among others, \cite{demauro}) , with the hope of a reform of school curricula that would reduce this sort of grammar-centrism \citep[cf., among others,][]{berretta}.
Results consistent with those presented here are, for example, those discussed in \cite{grassi}, \cite{desantis} and, before that, in \cite{lavinio}.

However, what emerges from this survey is the link between the two skills. In fact, in the  extension with covariates —which does not alter the latent structure but enhances interpretability by uncovering educationally meaningful clusters, offering valuable insights for targeted interventions in academic writing development— combinations of the two skills have been identified: Group 1 performs well in both linguistic and communicative competences, Group 2 shows low linguistic competence but high communicative competence, and Group 3 scores low in both competences. There does not seem to be a group with solid linguistic and grammatical competence but weak communicative competence. It therefore appears that grammatical competence implies communicative competence, but not vice versa. Thus, it is possible to have good communicative competence without equally solid grammatical competence. This consideration leads us to confirm what has been repeatedly stated in the literature regarding the weakness of a language education system based on an excessive focus on grammar, to the detriment of real communication and usage: there is still a certain gap between the language taught at school (very close to the old standard) and that used in moderately formal situations in everyday life. In this scenario, the language of university students, as seen in Figure \ref{figuova} and Figure \ref{figcovuni}, does not appear as a monolithic block, but varies according to social conditioning; in general terms, we can conclude that the language used by university students does not faithfully reproduce either of these standards, but represents a sort of compromise between them, consistent with the fact that university students combine the innovative drive typical of younger generations with the conservative tendencies typical of advanced education.

Beyond the present case study, the proposed framework is applicable to a broader class of problems involving multivariate count outcomes with complex dependence structures and latent heterogeneity, including textual data, educational assessments, psychological and behavioral studies, and other count settings.

\clearpage
\begin{appendices}

\counterwithin{figure}{section}
\counterwithin{table}{section}

\renewcommand{\thefigure}{\thesection\arabic{figure}}
\renewcommand{\thetable}{\thesection\arabic{table}}

\titleformat{\section} 
  {\normalfont\Large\bfseries} 
  {Appendix \thesection:}       
  {0.4em}                          
  {}

\section{Data}\label{sec_app_data}

\begin{figure}[h]
\begin{center}
\includegraphics[width=0.68\textwidth]{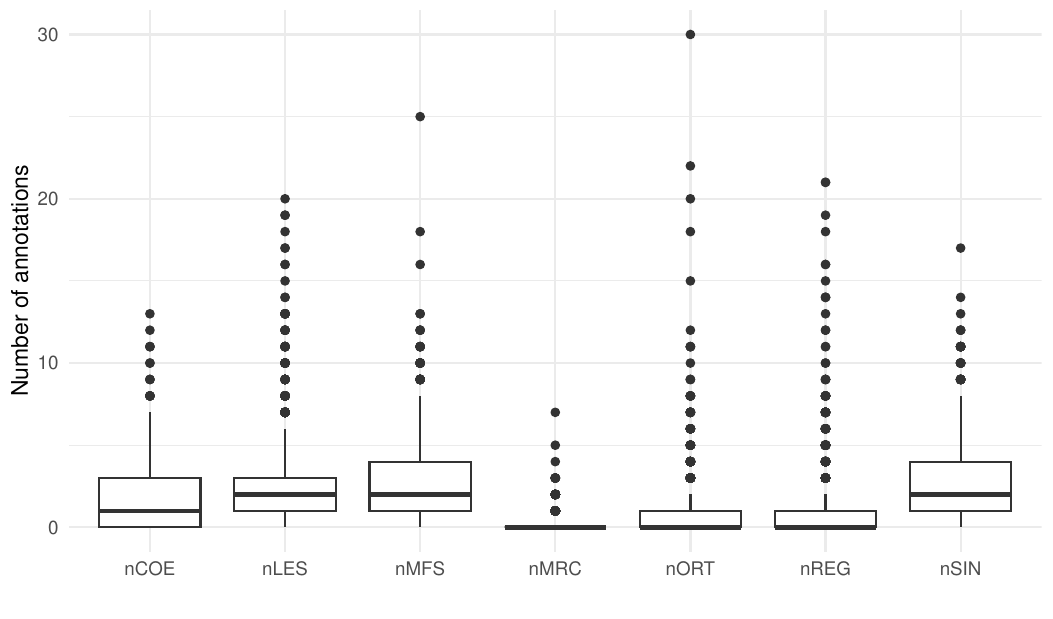}
\end{center}
\caption{Boxplots representing the distribution of the number of annotations for each feature considered in the analysis.}
\label{boxplot_annotations}
\end{figure}

\section{Generalized EM algorithm}\label{sec11}

\subsection*{E-step}\label{subsec4}
To calculate the expected value in Equation \ref{eqem}, $f(\mathbf{z},\mathbf{s}|\mathbf{y},\boldsymbol{\theta}')$ needs to be computed. The latter can be written as follows:
\begin{equation*}
f(\mathbf{z},\mathbf{s}|\mathbf{y};\boldsymbol{\theta}')=f(\mathbf{s}|\mathbf{y};\boldsymbol{\theta}')f(\mathbf{z}|\mathbf{y},\mathbf{s};\boldsymbol{\theta}'). 
\end{equation*}
According to the Bayes' rule $f(\mathbf{z}|\mathbf{y},s_i=1;\boldsymbol{\theta}')$ can be decomposed as:
$$f(\mathbf{z}|\mathbf{y},s_i=1;\boldsymbol{\theta}')=\dfrac{f(\mathbf{y}|\mathbf{z};\boldsymbol{\theta}')f(\mathbf{z}|s_i=1;\boldsymbol{\theta}')}{f(\mathbf{y}|s_i=1;\boldsymbol{\theta}')}. $$

$f(\mathbf{z}|s_i=1)$ follows a multivariate Gaussian density with as mean the vector $\boldsymbol{\mu}_i$ and as covariance the matrix $\boldsymbol{\Sigma}_i$, while $f(\mathbf{y}|\mathbf{z})$ is provided in Equation \ref{obslay}.

On the contrary, the denominator needs to be approximated because it cannot be written in closed form. Approximating it using Gauss-Hermite quadrature points we obtain:
\begin{equation*}
f(\mathbf{y}|s_i=1;\boldsymbol{\theta}')=\int f(\mathbf{y}|\mathbf{z};\boldsymbol{\theta}')f(\mathbf{z}|s_i=1;\boldsymbol{\theta}')d\mathbf{z} \approx \sum_{t_1=1}^{T_1}\ldots\sum_{t_q=1}^{T_q} w_{t_1}\ldots w_{t_q} f(\mathbf{y}|\sqrt{2}\boldsymbol{\Sigma}_i^{1/2}\mathbf{z}_t+\boldsymbol{\mu}_i;\boldsymbol{\theta}'),
\end{equation*}
where $w_{t_1},\ldots,w_{t_q}$ are the weights and $\mathbf{z}_t=(z_{t_1},\ldots,z_{t_q})^T$ the points of the quadrature. 

The Bayes' rule can be involved again to compute $f(\mathbf{s}|\mathbf{y};\boldsymbol{\theta}')$:
\begin{equation*}
f(s_i=1|\mathbf{y};\boldsymbol{\theta}')=\dfrac{f(\mathbf{y}|s_i=1;\boldsymbol{\theta}')f(s_i=1;\boldsymbol{\theta}')}{\sum_{i=1}^kf(\mathbf{y}|s_i=1;\boldsymbol{\theta}')f(s_i=1;\boldsymbol{\theta}')}.
\end{equation*}


\subsection*{M-step}\label{subsec5}

Let's start by finding $\tilde{\boldsymbol{\Lambda}}$ maximizing $E_{\mathbf{z},\mathbf{s}|\mathbf{y},\boldsymbol{\theta'}}[log f(\mathbf{y}|\mathbf{z};\mathbf{\boldsymbol{\theta}})]$. The expected value is:
\begin{equation}
E_{\mathbf{z},\mathbf{s}|\mathbf{y},\boldsymbol{\theta'}}[log f(\mathbf{y}|\mathbf{z};\mathbf{\boldsymbol{\theta}})]=E_{\mathbf{z}|\mathbf{y},\boldsymbol{\theta'}}[log f(\mathbf{y}|\mathbf{z};\mathbf{\boldsymbol{\theta}})]= \int log\,f(\mathbf{y}|\mathbf{z};\mathbf{\boldsymbol{\theta}}) f(\mathbf{z}|\mathbf{y};\boldsymbol{\theta}')d\mathbf{z},
\label{eqmstep1}
\end{equation}
with 
$$f(\mathbf{z}|\mathbf{y};\boldsymbol{\theta}')=\sum_{i=1}^k f(s_i=1|\mathbf{y};\boldsymbol{\theta}')f(\mathbf{z}|\mathbf{y},s_i=1; \boldsymbol{\theta}'). $$ 


Equation \ref{eqmstep1} needs to be maximized with respect to $\tilde{\boldsymbol{\lambda}}_j=(\lambda_{j0},\lambda_{j1},\ldots,\lambda_{jq})$, for $j=1,\ldots,p$. In order to do that, the derivative of $log f(\mathbf{y}|\mathbf{z};\mathbf{\boldsymbol{\theta}})$ must be computed. Since $log f(\mathbf{y}|\mathbf{z};\mathbf{\boldsymbol{\theta}})=log f(\mathbf{y}|\tilde{\mathbf{z}};\mathbf{\boldsymbol{\theta}})$, where $\tilde{\mathbf{z}}=(1,\mathbf{z})$, we get:
\begin{equation*}
\begin{split}
S_0(\tilde{\boldsymbol{\lambda}}_j,\mathbf{y}|\tilde{\mathbf{z}})&=\dfrac{d\,log f(\mathbf{y}|\tilde{\mathbf{z}};\mathbf{\boldsymbol{\theta}})}{d\tilde{\boldsymbol{\lambda}}_j}=\dfrac{d}{d\tilde{\boldsymbol{\lambda}}_j}[y_j\, log(\omega_j)-\omega_j-log(y_j!)]=\\
&\\
&=\dfrac{d}{d\tilde{\boldsymbol{\lambda}}_j} [y_j\,(\lambda_{j0}+\boldsymbol{\lambda}_j^T\mathbf{z})-exp(\lambda_{j0}+\boldsymbol{\lambda}_j^T \mathbf{z})-log(y_j!)]. \\
\end{split}
\end{equation*}
The last equality follows from the fact that $\omega_j=exp(\lambda_{j0}+\boldsymbol{\lambda}_j^T \mathbf{z})$. 
Thus:
$$S_0(\tilde{\boldsymbol{\lambda}}_j,\mathbf{y}|\tilde{\mathbf{z}})=[y_j \psi_j'-b_j'(\psi_j)], $$
where $\psi_j$ is the canonical parameter. 
\\Deriving we get:
$$
\psi_j'=
\left \{
\begin{array}{l}
1 \hspace{0.5cm}\text{if}\hspace{0.5cm} \tilde{\boldsymbol{\lambda}}_j=\lambda_{j0}; \\
\mathbf{z} \hspace{0.5cm}\text{if}\hspace{0.5cm} \tilde{\boldsymbol{\lambda}}_j=\boldsymbol{\lambda}_j.\\

\end{array}
\right.
$$
\\
$$
b_j'(\psi_j)=
\left \{
\begin{array}{l}
exp(\lambda_{j0}+\boldsymbol{\lambda}_j^T\mathbf{z})=\omega_j \hspace{0.935cm}\text{if}\hspace{0.5cm} \tilde{\boldsymbol{\lambda}}_j=\lambda_{j0}; \\
\mathbf{z}\,exp(\lambda_{j0}+\boldsymbol{\lambda}_j^T\mathbf{z})=\mathbf{z}\,\omega_j \hspace{0.5cm}\text{if}\hspace{0.5cm} \tilde{\boldsymbol{\lambda}}_j=\boldsymbol{\lambda}_j.\\

\end{array}
\right.
$$
Thus, rewriting Equation \ref{eqmstep1} as:
\begin{equation*}
\begin{split}
E_{\mathbf{z}|\mathbf{y},\boldsymbol{\theta'}}[log f(\mathbf{y}|\mathbf{z};\mathbf{\boldsymbol{\theta}})]&= \int log[f(\mathbf{y}|\mathbf{z};\mathbf{\boldsymbol{\theta}})] f(\mathbf{z}|\mathbf{y};\boldsymbol{\theta}')d\mathbf{z}=\\
&=\int log[f(\mathbf{y}|\mathbf{z};\mathbf{\boldsymbol{\theta}})]\sum_{i=1}^k f(s_i=1|\mathbf{y};\boldsymbol{\theta}')f(\mathbf{z}|\mathbf{y},s_i=1; \boldsymbol{\theta}') d\mathbf{z}, \\
\end{split}
\end{equation*}
the expected score function with respect to $\tilde{\boldsymbol{\lambda}}_j$ becomes:
$$\dfrac{d\,E_{\tilde{\mathbf{z}}|\mathbf{y},\boldsymbol{\theta'}}log f(\mathbf{y}|\tilde{\mathbf{z}};\mathbf{\boldsymbol{\theta}})}{d\tilde{\boldsymbol{\lambda}}_j}=0 $$
$$\sum_{i=1}^k f(s_i=1|\mathbf{y};\boldsymbol{\theta}') \int S_0(\tilde{\boldsymbol{\lambda}}_j, \mathbf{y}|\tilde{\mathbf{z}}) f(\tilde{\mathbf{z}}|\mathbf{y},s_i=1; \boldsymbol{\theta}') d\tilde{\mathbf{z}}=0 $$
$$\sum_{i=1}^k f(s_i=1|\mathbf{y};\boldsymbol{\theta}') \bigintssss S_0(\tilde{\boldsymbol{\lambda}}_j, \mathbf{y}|\tilde{\mathbf{z}}) f(\tilde{\mathbf{z}}|s_i=1; \boldsymbol{\theta}')\dfrac{f(\mathbf{y}|\tilde{\mathbf{z}}; \boldsymbol{\theta}')}{f(\mathbf{y}|s_i=1;\boldsymbol{\theta}')} d\tilde{\mathbf{z}}=0 $$

The integral can be approximated using Gauss-Hermite quadrature points, since $f(\tilde{\mathbf{z}}|s_i=1; \tilde{\boldsymbol{\theta}}')$ follows a multivariate Gaussian density. Thus:
$$
\bigintssss S_0(\tilde{\boldsymbol{\lambda}}_j, \mathbf{y}|\tilde{\mathbf{z}}) f(\tilde{\mathbf{z}}|s_i=1; \boldsymbol{\theta}')\dfrac{f(\mathbf{y}|\tilde{\mathbf{z}}; \boldsymbol{\theta}')}{f(\mathbf{y}|s_i=1;\boldsymbol{\theta}')} d\tilde{\mathbf{z}}\simeq$$
$$\simeq \,\dfrac{1}{f(\mathbf{y}|s_i=1;\boldsymbol{\theta}')}\sum_{t_1=1}^{T_1}\ldots\sum_{t_q=1}^{T_q} w_{t_1}\ldots w_{t_q} S_0(\tilde{\boldsymbol{\lambda}}_j,\mathbf{y}|\tilde{\mathbf{z}}_t^*)f(\mathbf{y}|\tilde{\mathbf{z}}_t^*;\boldsymbol{\theta}'),  $$
with $\tilde{\mathbf{z}}_t^*=(1,\mathbf{z}_t^*)$ and $\mathbf{z}_t^*=\sqrt{2}\,\boldsymbol{\Sigma}_i^{1/2}\mathbf{z}_t+\boldsymbol{\mu}_i$.

However, this approximate gradient does not provide an explicit solution and, in order to find the estimate of $\tilde{\boldsymbol{\Lambda}}$, nonlinear optimization methods need to be employed. In this work Newton-type algorithms \citep{newton} have been adopted. \\

To estimate $\boldsymbol{\mu}$ and $\boldsymbol{\Sigma}$, $E_{\mathbf{z},\mathbf{s}|\mathbf{y},\boldsymbol{\theta}'} [log\, f(\mathbf{z}|\mathbf{s};\boldsymbol{\theta})]$ needs to be maximized with respect to the two parameters of interest.
The expected value is:
\begin{equation*}
\begin{split}
E_{\mathbf{z},\mathbf{s}|\mathbf{y},\boldsymbol{\theta}'} [log\, f(\mathbf{z}|\mathbf{s};\boldsymbol{\theta})]&= 
\sum_{i=1}^k \int log[\, f(\mathbf{z}|s_i=1;\boldsymbol{\theta})] f(\mathbf{z},s_i=1|\mathbf{y},\boldsymbol{\theta}') d\mathbf{z}\\
&=\sum_{i=1}^k \int log[\, f(\mathbf{z}|s_i=1;\boldsymbol{\theta})] f(\mathbf{z}|\mathbf{y},s_i=1;\boldsymbol{\theta}') f(s_i=1|\mathbf{y};\boldsymbol{\theta}') d\mathbf{z}\\
&=\sum_{i=1}^k f(s_i=1|\mathbf{y};\boldsymbol{\theta}') \int log[\, f(\mathbf{z}|s_i=1;\boldsymbol{\theta})] f(\mathbf{z}|\mathbf{y},s_i=1;\boldsymbol{\theta}') d\mathbf{z}\\
\end{split}
\end{equation*}
$$=\sum_{i=1}^k f(s_i=1|\mathbf{y};\boldsymbol{\theta}') \bigintsss log\Biggl[\dfrac{det(\boldsymbol{\Sigma}_i)^{-1/2}}{(2\pi)^{q/2}}exp\Biggl\{-\dfrac{(\mathbf{z}-\boldsymbol{\mu}_i)^T\boldsymbol{\Sigma}_i^{-1}(\mathbf{z}-\boldsymbol{\mu}_i)}{2}\Biggr\}\Biggr] f(\mathbf{z}|\mathbf{y},s_i=1;\boldsymbol{\theta}')d\mathbf{z}$$
$$=\sum_{i=1}^k f(s_i=1|\mathbf{y};\boldsymbol{\theta}') \bigintsss \Biggl[-\dfrac{q}{2}log(2\pi)-\dfrac{1}{2}log(det(\boldsymbol{\Sigma}_i))-\dfrac{(\mathbf{z}-\boldsymbol{\mu}_i)^T\boldsymbol{\Sigma}_i^{-1}(\mathbf{z}-\boldsymbol{\mu}_i)}{2}\Biggr] f(\mathbf{z}|\mathbf{y},s_i=1;\boldsymbol{\theta}')d\mathbf{z}$$
Maximizing this expectation with respect to $\boldsymbol{\mu}_i$, the corresponding estimate is obtained:
\begin{equation}
\boldsymbol{\mu}_i=\dfrac{E(\mathbf{z}|\mathbf{y},s_i=1;\boldsymbol{\theta}')f(s_i=1|\mathbf{y};\boldsymbol{\theta}') }{f(s_i=1|\mathbf{y};\boldsymbol{\theta}') }, 
\end{equation}

where $E(\mathbf{z}|\mathbf{y},s_i=1;\boldsymbol{\theta}')$ can be approximated using Gauss-Hermite quadrature points as:
\begin{equation*}
\begin{split}
E(\mathbf{z}|\mathbf{y},s_i=1;\boldsymbol{\theta}')&=\int \mathbf{z} f(\mathbf{z}|s_i=1,\mathbf{y};\boldsymbol{\theta}')d\mathbf{z}=\bigintssss \mathbf{z}\,\dfrac{f(\mathbf{y}|\mathbf{z};\boldsymbol{\theta}')f(\mathbf{z}|s_i=1;\boldsymbol{\theta}')}{f(\mathbf{y}|s_i=1;\boldsymbol{\theta}')}d\mathbf{z} \\
&=\dfrac{1}{f(\mathbf{y}|s_i=1;\boldsymbol{\theta}')}\int  \mathbf{z}f(\mathbf{y}|\mathbf{z};\boldsymbol{\theta}')f(\mathbf{z}|s_i=1;\boldsymbol{\theta}')d\mathbf{z}
\end{split}
\end{equation*}
$f(\mathbf{z}|s_i=1;\boldsymbol{\theta}')$ follows a multivariate Gaussian distribution, thus:
\begin{equation*}
E(\mathbf{z}|\mathbf{y},s_i=1;\boldsymbol{\theta}')=\dfrac{1}{f(\mathbf{y}|s_i=1;\boldsymbol{\theta}')} \sum_{t_1=1}^{T_1}\ldots\sum_{t_q=1}^{T_q}\omega_{t_1}\ldots\omega_{t_q} \mathbf{z}_t^*f(\mathbf{y}|\mathbf{z}_t^*;\boldsymbol{\theta}').
\end{equation*}

To get an estimate of $\boldsymbol{\Sigma}_i$, the same expected value is maximized with respect to the covariance matrix. It results that:

\begin{equation}
\boldsymbol{\Sigma}_i=\dfrac{f(s_i=1|\mathbf{y};\boldsymbol{\theta}')[E(\mathbf{z}\mathbf{z}^T|\mathbf{y},s_i=1;\boldsymbol{\theta}')-\boldsymbol{\mu}_i \boldsymbol{\mu}_i^T]}{f(s_i=1|\mathbf{y};\boldsymbol{\theta}')}. 
\end{equation}

$E(\mathbf{z}\mathbf{z}^T|\mathbf{y},s_i=1;\boldsymbol{\theta}')$ admits the following approximation based on Gauss-Hermite quadrature points:
\begin{equation*}
\begin{split}
E(\mathbf{z}\mathbf{z}^T|\mathbf{y},s_i=1;\boldsymbol{\theta}')&=\int \mathbf{z}\mathbf{z}^Tf(\mathbf{z}|\mathbf{y},s_i=1;\boldsymbol{\theta}')d\mathbf{z} \\
&=\dfrac{1}{f(\mathbf{y}|s_i=1;\boldsymbol{\theta}')}\int \mathbf{z}\mathbf{z}^Tf(\mathbf{y}|\mathbf{z};\boldsymbol{\theta}')f(\mathbf{z}|s_i=1;\boldsymbol{\theta}')d\mathbf{z} \\
&=\dfrac{1}{f(\mathbf{y}|s_i=1;\boldsymbol{\theta}')} \sum_{t_1=1}^{T_1}\ldots\sum_{t_q=1}^{T_q} \omega_{t_1}\ldots \omega_{t_q}\, \mathbf{z}_t^{*}\mathbf{z}_t^{*T}f(\mathbf{y}|\mathbf{z}_t^*;\boldsymbol{\theta}').
\end{split}
\end{equation*}

To account for the identifiability conditions mentioned before, at each iteration of the EM algorithm the centering and scaling transformations below are adopted:
\begin{equation*}
\begin{split}
&\boldsymbol{\Sigma}_i \rightarrow (\mathbf{A}^{-1})^T\boldsymbol{\Sigma}_i\,\mathbf{A}^{-1}\\
&\boldsymbol{\mu}_i \rightarrow (\mathbf{A}^{-1})^T\boldsymbol{\mu}_i\\
&\boldsymbol{\mu}_i \rightarrow \boldsymbol{\mu}_i-\sum_{i=1}^k \pi_i\, \boldsymbol{\mu}_i\\
\end{split}
\end{equation*}
where $\mathbf{A}$ represents the Cholesky decomposition of the variance of $\mathbf{z}$.

Finally, the mixture weights $\boldsymbol{\pi}$ can be estimated maximizing the subsequent expected value: 
\begin{equation*}
\begin{split}
E_{\mathbf{z},\mathbf{s}|\mathbf{y},\boldsymbol{\theta'}} [log f(\mathbf{s;\boldsymbol{\theta}})]&=\sum_{i=1}^k\int log\,f(s_i=1;\boldsymbol{\theta})f(\mathbf{z},s_i=1|\mathbf{y}, \boldsymbol{\theta}')d\mathbf{z}\\ 
&=\sum_{i=1}^k \int (s_i=1) log(\pi_i) f(s_i=1|\mathbf{y};\boldsymbol{\theta}')f(\mathbf{z}|\mathbf{y},s_i=1;\boldsymbol{\theta}')d\mathbf{z}\\
&=\sum_{i=1}^k  (s_i=1) log(\pi_i) f(s_i=1|\mathbf{y};\boldsymbol{\theta}').
\end{split}
\end{equation*}

However, this optimization does not return an explicit expression for the mixture weights. Thus, Lagrange multipliers are involved. 
The final estimate of the mixture weight $\pi_i$ is:
\begin{equation}
\pi_i=f(s_i=1|\mathbf{y};\boldsymbol{\theta}').
\end{equation}

\section{Simulation Study}\label{sec12}

\begin{table}[H]
\caption{Model selection results for the additional simulation designs considered. For each setting, bold identifies the true model specification and the corresponding percentage of times such model was selected by each criterion.}
\small
\setlength{\tabcolsep}{3pt}
\begin{center}
\begin{tabular}{lccc|ccc|lccc|ccc}

\hline
& \multicolumn{3}{c}{BIC}& \multicolumn{3}{c|}{AIC} & & \multicolumn{3}{c}{BIC}& \multicolumn{3}{c}{AIC} \\
\hline
 &\multicolumn{6}{c|}{$k=2, q=1, n=300, p=50$}&& \multicolumn{6}{c}{$k=2, q=2, n=300,p=50$}\\
& $\boldsymbol{q=1}$ &  $q=2$  & $q=3$ 
& $\boldsymbol{q=1}$ &  $q=2$  & $q=3$  && $q=1$ &  $\boldsymbol{q=2}$  & $q=3$
& $q=1$ &  $\boldsymbol{q=2}$  & $q=3$\\
\hline

$k$=1 &0\% &0\%&0\% &0\%&0\%&0\% & $k$=1 &0\% &0\%&0\% &0\%&0\%&0\%\\
$\boldsymbol{k}$=\textbf{2} &$\mathbf{45}$\% &0\% & 0\%&$\mathbf{40}$\%&4\%&0\%&
$\boldsymbol{k}$=\textbf{2} &0\% &$\mathbf{55}$\% & 13\%&0\%&$\mathbf{1}$\%&54\%\\
$k$=3 &55\% &0\%&0\%&53\%&1\%&0\%&$k$=3 &0\% &28\%&0\%&0\%&1\%&22\%\\
$k$=4 &0\% &0\%&0\%&2\%&0\%&0\%& $k$=4 &0\% &4\%&0\%&0\%&0\%&22\%\\
\hline

%
%
&\multicolumn{6}{c|}{$k=2, q=1, n=2000,p=50$}&&\multicolumn{6}{c}{$k=2, q=2, n=2000,p=50$}\\

& $\boldsymbol{q=1}$ &  $q=2$  & $q=3$
& $\boldsymbol{q=1}$ &  $q=2$  & $q=3$  && $q=1$ &  $\boldsymbol{q=2}$  & $q=3$
& $q=1$ &  $\boldsymbol{q=2}$  & $q=3$\\
\hline
$k$=1 &0\% &0\%&0\% &0\%&0\%&0\%&$k$=1 &0\% &0\%&0\% &0\%&0\%&0\%\\
$\boldsymbol{k}$=\textbf{2} &$\mathbf{71}$\% &0\% & 0\%&$\mathbf{59}$\%&3\%&0\%&
$\boldsymbol{k}$=\textbf{2} &0\% &$\mathbf{0}$\% & 100\%&0\%&$\mathbf{0}$\%&19\%\\
$k$=3 &29\% &0\%&0\%&37\%&0\%&0\%& $k$=3 &0\% &0\%&0\%&0\%&0\%&44\%\\
$k$=4 &0\% &0\%&0\%&1\%&0\%&0\%& $k$=4 &0\% &0\%&0\%&0\%&0\%&37\%\\

\hline

&\multicolumn{6}{c|}{$k=3, q=1, n=300,p=50$}&&\multicolumn{6}{c}{$k=3, q=2, n=300,p=50$}\\

& $\boldsymbol{q=1}$ &  $q=2$  & $q=3$
& $\boldsymbol{q=1}$ &  $q=2$  & $q=3$ && $q=1$ &  $\boldsymbol{q=2}$  & $q=3$
& $q=1$ &  $\boldsymbol{q=2}$  & $q=3$  \\
\hline
$k$=1 &0\% &0\%&0\% &0\%&0\%&0\%&$k$=1 &0\% &0\%&0\% &0\%&0\%&0\%\\
$k$=2 &0\% &0\% & 0\%&0\%&0\%&0\%&$k$=2 &0\% &0\% & 0\%&0\%&0\%&0\%\\
$\boldsymbol{k}$=\textbf{3} &$\mathbf{80}$\% &0\%&0\%&$\mathbf{79}$\%&2\%&0\%&$\boldsymbol{k}$=\textbf{3} &0\% &$\mathbf{98}$\%&0\%&0\%&$\mathbf{88}$\%&3\%\\
$k$=4 &20\% &0\%&0\%&19\%&0\%&0\%&$k$=4 &0\% &2\%&0\%&0\%&8\%&1\%\\

\hline

%
%

&\multicolumn{6}{c|}{$k=3, q=1, n=2000,p=50$}&& \multicolumn{6}{c}{$k=3, q=2, n=2000,p=50$}\\

& $\boldsymbol{q=1}$ &  $q=2$  & $q=3$
& $\boldsymbol{q=1}$ &  $q=2$  & $q=3$  && $q=1$ &  $\boldsymbol{q=2}$  & $q=3$
& $q=1$ &  $\boldsymbol{q=2}$  & $q=3$  \\
\hline
$k$=1 &0\% &0\%&0\% &0\%&0\%&0\%& $k$=1 &0\% &0\%&0\% &0\%&0\%&0\%\\
$k$=2 &0\% &0\% & 0\%&0\%&0\%&0\%&  $k$=2 &0\% &0\% & 0\%&0\%&0\%&0\%\\
$\boldsymbol{k}$=\textbf{3} &$\mathbf{90}$\% &0\%&0\%&$\mathbf{90}$\%&0\%&0\% & $\boldsymbol{k}$=\textbf{3} &0\% &$\mathbf{100}$\%&0\%&0\%&$\mathbf{69}$\%&1\%\\
$k$=4 &10\% &0\%&0\%&10\%&0\%&0\%& $k$=4 &0\% &0\%&0\%&0\%&30\%&0\%\\

\hline
&\multicolumn{6}{c|}{$k=6, q=1, n=300,p=50$}&&\multicolumn{6}{c}{$k=6, q=2, n=300,p=50$}\\

& $\boldsymbol{q=1}$ &  $q=2$  & $q=3$
& $\boldsymbol{q=1}$ &  $q=2$  & $q=3$ &&  $q=1$ &  $\boldsymbol{q=2}$  & $q=3$
& $q=1$ &  $\boldsymbol{q=2}$  & $q=3$   \\
\hline
$k$=1 &0\% &0\%&0\% &0\%&0\%&0\%&$k$=1 &0\% &0\%&0\% &0\%&0\%&0\%\\
$k$=2 &0\% &0\% & 0\%&0\%&1\%&0\%&$k$=2 &0\% &0\% & 0\%&0\%&0\%&0\%\\
$k$=3 &22\% &0\%&0\%&2\%&7\%&0\%&$k$=3 &0\% &49\%&0\%&0\%&0\%&0\%\\
$k$=4 &32\% &0\%&0\%&1\%&39\%&1\%&$k$=4 &0\% &43\%&0\%&0\%&4\%&0\%\\
$k$=5 &33\% &0\%&0\%&1\%&33\%&0\%&$k$=5 &0\% &8\%&0\%&0\%&12\%&12\%\\
$\boldsymbol{k}$=\textbf{6} &$\mathbf{12}$\% &0\%&0\%&$\mathbf{3}$\%&6\%&0\%&$\boldsymbol{k}$=\textbf{6} &0\% &$\mathbf{0}$\%&0\%&0\%&$\mathbf{50}$\%&5\%\\
$k$=7 &1\% &0\%&0\%&2\%&4\%&0\%&$k$=7 &0\% &0\%&0\%&0\%&17\%&0\%\\

\hline

%
%
&\multicolumn{6}{c|}{$k=6, q=1, n=2000,p=50$}&&\multicolumn{6}{c}{$k=6, q=2, n=2000,p=50$}\\

& $\boldsymbol{q=1}$ &  $q=2$  & $q=3$
& $\boldsymbol{q=1}$ &  $q=2$  & $q=3$  && $q=1$ &  $\boldsymbol{q=2}$  & $q=3$
& $q=1$ &  $\boldsymbol{q=2}$  & $q=3$ \\
\hline
$k$=1 &0\% &0\%&0\% &0\%&0\%&0\%&$k$=1 &0\% &0\%&0\% &0\%&0\%&0\%\\
$k$=2 &0\% &0\% & 0\%&0\%&0\%&0\%&$k$=2 &0\% &0\% & 0\%&0\%&0\%&0\%\\
$k$=3 &1\% &0\%&0\%&0\%&0\%&0\%&$k$=3 &0\% &0\%&0\%&0\%&0\%&0\%\\
$k$=4 &3\% &34\%&0\%&0\%&23\%&0\%&$k$=4 &0\% &4\%&0\%&0\%&0\%&0\%\\
$k$=5 &28\% &15\%&0\%&0\%&19\%&9\%&$k$=5 &0\% &4\%&0\%&0\%&1\%&3\%\\
$\boldsymbol{k}$=\textbf{6} &$\mathbf{3}$\% &5\%&0\%&$\mathbf{0}$\%&19\%&7\%&$\boldsymbol{k}$=\textbf{6} &0\% &$\mathbf{73}$\%&0\%&0\%&$\mathbf{61}$\%&4\%\\
$k$=7 &11\% &0\%&0\%&0\%&17\%&6\%&$k$=7 &0\% &19\%&0\%&0\%&29\%&2\%\\

\hline

\end{tabular}
\end{center}
\label{table1gllvm_appendix}
\end{table}

\begin{table}[h!]
\small
\setlength{\tabcolsep}{3pt}
\caption{Performance of the models selected by BIC and AIC in recovering the true clustering structure. For each criterion, the left block reports the median misclassification rate and, in brackets, the first and third quartiles; the right block reports the median adjusted Rand Index and, in brackets, the corresponding first and third quartiles, computed over the 100 replicates.}
     \begin{center}
        \begin{tabular}{llcc|cc}
\hline
\multicolumn{2}{c}{}& \multicolumn{2}{c|}{BIC}& \multicolumn{2}{c}{AIC}\\
\multicolumn{2}{c}{}& \multicolumn{1}{c}{MISC}& \multicolumn{1}{c|}{aRI}&\multicolumn{1}{c}{MISC}& \multicolumn{1}{c}{aRI}\\
\hline

\multirow{1}{*}{$k=2$}
& $n=300, p=50$ &0.007 [0.003;0.025]  &0.973[0.941;0.987]  &0.007[0.003;0.030]&0.973[0.940;0.987]\\
$q=1$& $n=2000, p=10$&0.032 [0.029;0.035]& 0.876[0.865;0.886]&0.032 [0.029;0.035] &0.876[0.865;0.886]  \\
 & $n=2000, p=50$ &0.004 [0.003;0.006]&0.982[0.976;0.988]&0.005[0.004;0.009]&0.982[0.972;0.986]  \\
\hline
\multirow{1}{*}{$k=2$}
& $n=300, p=50$& 0.010[0.007;0.072] &0.960 [0.867;0.973] &0.017[0.007;0.111]&0.934[0.792;0.973]\\
$q=2$& $n=2000, p=10$& 0.034 [0.032;0.036] &0.868[0.861;0.876]&0.034[0.032;0.036]&0.867[0.859;0.875]  \\
& $n=2000, p=50$& 0.006 [0.004;0.007] &0.976[0.972;0.982]&0.072[0.025;0.138]&0.858[0.782;0.942]  \\

\hline
\multirow{1}{*}{$k=3 $}
& $n=300, p=50$& 0.000 [0.000;0.000] &1.000[1.000;1.000]   &0.000[0.000;0.000] &1.000[1.000;1.000]\\
$q=1$& $n=2000, p=10$ &0.079 [0.074;0.083] &0.787[0.779;0.799]&0.079[0.074;0.083]  &0.788[0.779;0.799]  \\
& $n=2000, p=50$& 0.000 [0.000;0.000] &1.000[0.998;1.000]& 0.000[0.000;0.000] &1.000[0.998;1.000]  \\
\hline
\multirow{1}{*}{$k=3$}
& $n=300, p=50$& 0.023 [0.017;0.030] &0.929[0.909;0.949]  &0.023[0.020;0.033]&0.929[0.904;0.944]\\
$q=2$& $n=2000, p=10$& 0.100 [0.092;0.101] &0.726[0.713;0.742]&0.100[0.092;0.105]&0.726[0.713;0.742]  \\
& $n=2000, p=50$ &0.021 [0.018;0.024] &0.937[0.930;0.945]&0.023[0.019;0.031]&0.935[0.921;0.942]  \\
\hline
\multirow{1}{*}{$k=6$}
& $n=300, p=50$& 0.520 [0.469;0.563] &0.347[0.303;0.416]  &0.317[0.220;0.350]&0.602[0.549;0.687]\\
$q=1$& $n=2000, p=10$& 0.534 [0.513;0.636] &0.345[0.255;0.367]&0.516[0.499;0.528]&0.362[0.349;0.373]  \\
& $n=2000, p=50$& 0.345 [0.321;0.460] &0.551[0.439;0.599] &0.216[0.202;0.321]&0.674[0.611;0.710]  \\
\hline
\multirow{1}{*}{$k=6$}
& $n=300, p=50$& 0.342 [0.303;0.457] &0.641[0.570;0.688]&  0.170[0.138;0.183]&0.778[0.757;0.807]\\
$q=2$& $n=2000, p=10$& 0.490 [0.480;0.495]& 0.502[0.494;0.510]&0.324[0.290;0.383]&0.545[0.526;0.568]  \\
& $n=2000, p=50$ &0.176 [0.163;0.185]& 0.787[0.774;0.799]& 0.173[0.156;0.182]&0.788[0.777;0.801]  \\
\hline
\end{tabular}
\end{center}
\label{tableariscelte}
\end{table}

\begin{figure}[H]
\begin{center}
\includegraphics[width=0.72\textwidth]{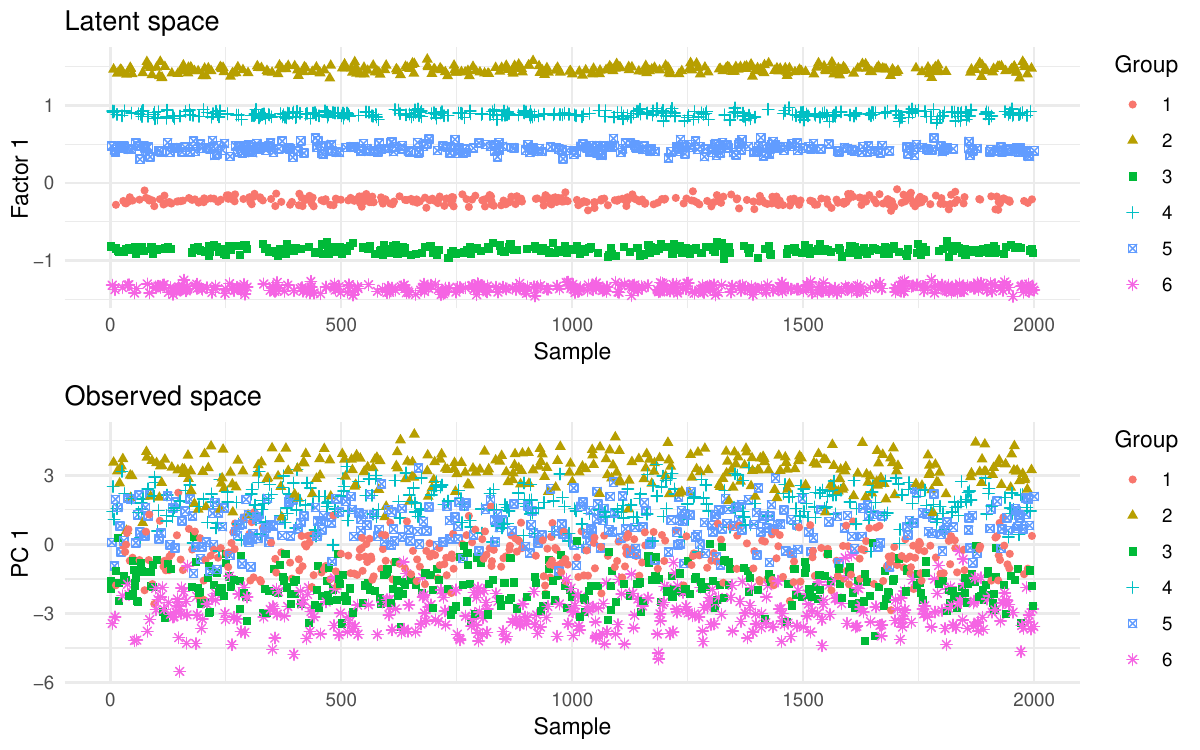}
\end{center}
\caption{Top: scatterplot of the factor scores distinguished by group for one of the 100 replicates of the experiment with $k=6, q=1, n=2000, p=10$;
Bottom: scatterplot of the first principal component (obtained with the \texttt{PLNPCA}  function \citep{PLNPCA} in the \texttt{R} \citep{Rcoreteam} package \texttt{PLNmodels} \citep{robin}) distinguished by group.}
\label{figpcak6q1}
\end{figure}

\begin{figure}[H]
\begin{center}
\includegraphics[width=0.72\textwidth]{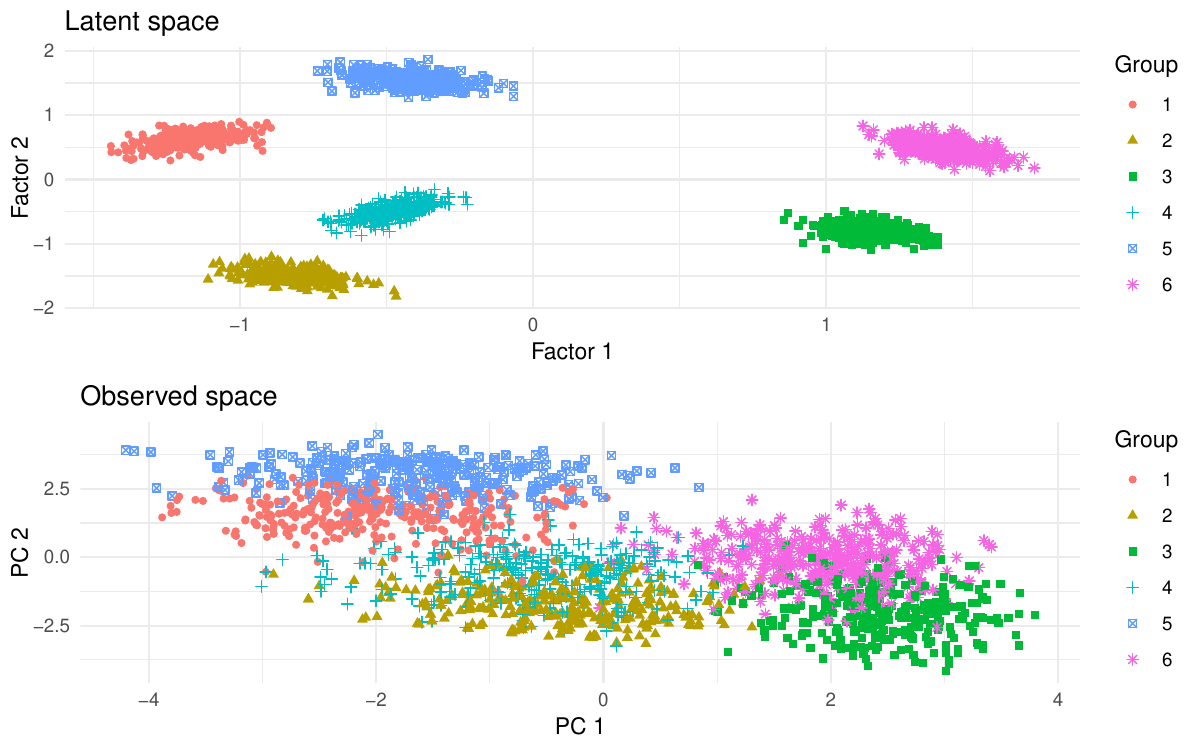}
\end{center}
\caption{Top: scatterplot of the factor scores distinguished by group for one of the 100 replicates of the experiment with $k=6, q=2, n=2000, p=10$;
Bottom: scatterplot of the first two principal components (obtained with the \texttt{PLNPCA} function \citep{PLNPCA} in the \texttt{R} \citep{Rcoreteam} package \texttt{PLNmodels} \citep{robin}) distinguished by group.}
\label{figpcak6q2}
\end{figure}

\end{appendices}

\clearpage

\section{Acknowledgments}
The research leading to these results has received funding from the European Union - NextGenerationEU through the Italian Ministry of University and Research under PNRR - M4C2-I1.3 Project PE\_00000019 “HEAL ITALIA” to Stefano Diciotti – CUP J33C22002920006. The views and opinions expressed are those of the authors only and do not necessarily reflect those of the European Union or the European Commission. Neither the European Union nor the European Commission can be held responsible for them.

\end{document}